\documentclass[lettersize,journal]{IEEEtran}
\usepackage{amsmath,amsfonts}
\usepackage{algorithmic}
\usepackage{algorithm}
\usepackage{array}
\usepackage{textcomp}
\usepackage{stfloats}
\usepackage{url}
\usepackage{verbatim}
\usepackage{graphicx}
\usepackage{multirow}
\usepackage{booktabs}
\usepackage{makecell}
\usepackage{xcolor}
\usepackage{hyperref}
\usepackage{subcaption}
\usepackage[numbers, sort&compress]{natbib}
\usepackage{bm}

\makeatletter
\newcommand\footnoteref[1]{\protected@xdef\@thefnmark{\ref{#1}}\@footnotemark}
\makeatother

\def\BibTeX{{\rm B\kern-.05em{\sc i\kern-.025em b}\kern-.08em
    T\kern-.1667em\lower.7ex\hbox{E}\kern-.125emX}}
\usepackage{balance}

\newcommand{\revision}[1]{\textcolor{black}{{#1}}}

\begin{document}

\title{Frequency-Aware Self-Supervised Music Representation Learning}

\author{Yicheng Gu, \textit{Student Member, IEEE}, Junan Zhang, Jerry Li, \\ Zhizheng Wu, \textit{Senior Member, IEEE}, Lauri Juvela, \textit{Member, IEEE}

\thanks{Jerry Li is with the Spellbrush, Akihabara, Tokyo 101-0021, Japan (e-mail: jerry@sizigistudios.com).}
\thanks{Lauri Juvela is with the Acoustic Lab, Department of Information and Communications Engineering (DICE), Aalto University, Espoo 02150, Finland (e-mail: lauri.juvela@aalto.fi).}
\thanks{Junan Zhang and Zhizheng Wu are with the School of Data Science, The Chinese University of Hong Kong, Shenzhen, Guangdong 518172, China (e-mail: junanzhang@link.cuhk.edu.cn;  wuzhizheng@cuhk.edu.cn).}
\thanks{Yicheng Gu is with the Spellbrush, Akihabara, Tokyo 101-0021, Japan, also with the Acoustic Lab, Department of Information and Communications Engineering (DICE), Aalto University, Espoo 02150, Finland, and also with the School of Data Science, The Chinese University of Hong Kong, Shenzhen, Guangdong 518172, China (e-mail:yichenggu@link.cuhk.edu.cn).}
}




\maketitle
 
\begin{abstract}


Self-supervised learning (SSL) has emerged as an essential paradigm for music information retrieval (MIR). 
While current SSL models achieve state-of-the-art performance across various MIR tasks, they typically treat audio as 1D sequences, either operating on time-domain waveforms or on flattened time-frequency-domain spectrograms.
This discards the rich spatial and structural information in time-frequency representations and overlooks a fundamental intuition in music production.
In particular, music is naturally represented as time-frequency grids in MIDI-based workflows, a structure that tightly corresponds to 2D spectrograms and inherently makes many MIR tasks trivial. 
\revision{Motivated by this intuition, we propose PupuM2D, a musical Masked Modeling Duo (M2D) model that is trained directly on 2D music spectrograms.}
\revision{Instead of applying masked language modeling (MLM) to 1D sequences, PupuM2D learns robust representations by predicting the latent embeddings of masked 2D music spectrogram patches from unmasked contexts.} To optimally adapt such a framework to music signals, we also apply domain-specific modifications to model architecture, training scheme, and inference paradigm, with comprehensive ablation studies showing their effectiveness.
\revision{Evaluations on the MARBLEv2 benchmark show that PupuM2D outperforms the sequence-based SSL models across multiple MIR tasks in linear probing.} 
\revision{Additionally, case studies of the attention maps also confirm that PupuM2D captures musically meaningful patterns within the 2D time-frequency domain. Codes and checkpoints are available at: \url{https://www.yichenggu.com/PupuM2D/}.}
\end{abstract}


\vspace{-10pt}

\section{Introduction}
\label{sec:introduction}

Music Information Retrieval (MIR) is a fundamental field encompassing a wide range of tasks, such as music tagging, genre classification, emotion recognition, and key detection. 
Traditionally, these tasks use supervised learning and require massive amounts of high-quality, human-annotated data~\cite{musiccnn}. 
As acquiring them is expensive and prone to subjective bias, the performance of such methods is often limited by the data bottlenecks. 
To solve this, self-supervised learning (SSL) has emerged as a mainstream paradigm in the MIR community. In particular, SSL models can efficiently learn universal acoustic and semantic representations from vast amounts of unlabeled audio, and can be easily adapted to various MIR tasks with competitive performances through simple linear probing.

In the early stages, music SSL models mainly relied on leveraging generative models or directly adapting existing frameworks from other domains. 
Specifically, JukeboxMIR~\cite{jukeboxmir} explored using the hidden states of Jukebox~\cite{jukebox} for MIR tasks; CLMR~\cite{clmr} adapted a contrastive learning framework with audio-specific data augmentations; and MULE~\cite{mule} further scaled this up by incorporating visual CNNs on spectrograms. 
However, their performance is often limited, as extracting features from generative models incurs high computational costs, and contrastive learning often relies on augmentations like cropping and rotation that are unnatural to music.

Recent studies shifted towards BERT-style masked language modeling (MLM)~\cite{bert} to address these limitations. In particular, MERT~\cite{mert} introduces a dual-teacher framework that leverages acoustic tokens from Encodec~\cite{encodec} and musical tokens from the Constant-Q Transform (CQT) to construct comprehensive targets that capture diverse aspects of audio; MusicFM~\cite{musicfm} bypasses external tokenizers by using random projections to generate masked targets; and MuQ~\cite{muq} proposes a lightweight Mel-RVQ tokenizer to generate robust quantization targets from spectrograms, which achieves state-of-the-art (SOTA) performance across diverse MIR benchmarks.

\begin{figure*}[t]
    \centering
    \includegraphics[width=\textwidth]{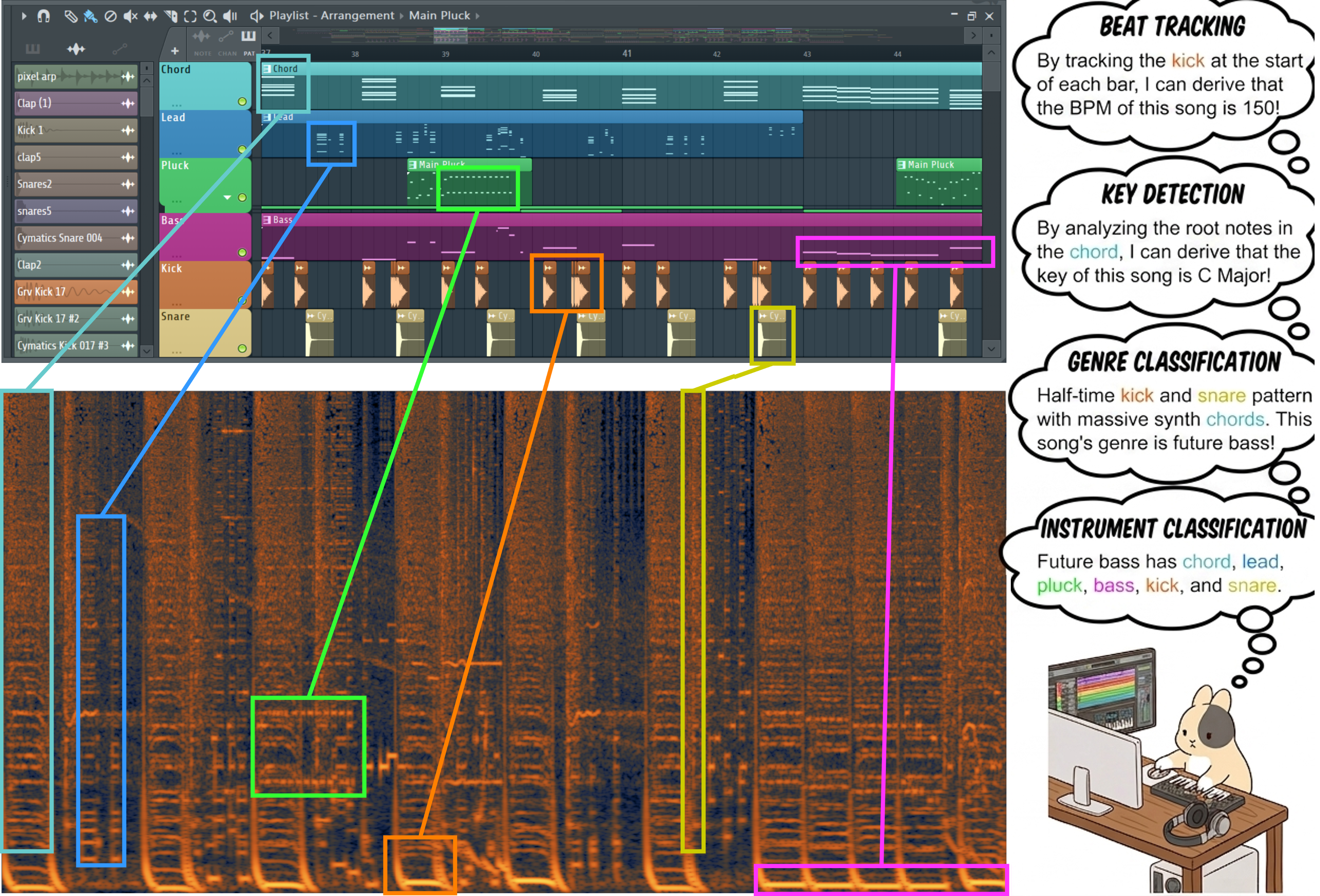}
    \caption{\revision{Illustration of the fundamental intuition behind PupuM2D}. The top panel displays a multitrack project in a Digital Audio Workstation (DAW), while the bottom panel shows its corresponding 2D spectrogram. Colored bounding boxes explicitly map individual tracks to their distinct time-frequency patterns in the spectrogram. As illustrated on the right, an experienced producer, Pupu (``bunny'' in Finnish), can intuitively perform MIR tasks by visually interpreting these 2D spectrogram patterns.}
    \label{fig:intuition}
    \vspace{-18pt}
\end{figure*}

However, a fundamental architectural bias persists across these recent works: they rigidly follow the model design in speech and natural language processing, treating audio as 1D sequences, either operating on time-domain waveforms~\cite{mert, musicfm} or on flattened time-frequency-domain spectrograms~\cite{muq}. Such an operation discards the rich 2D spatial and harmonic structures preserved in time-frequency representations, thereby restricting the models' ability for fine-grained and localized musical analysis. This reduction also overlooks a fundamental intuition in modern music production. As shown in Figure~\ref{fig:intuition}, in MIDI-based workflows, music is naturally represented as time-frequency grids, a structure that tightly corresponds to 2D spectrograms and makes many MIR tasks trivial.

\revision{Motivated by this intuition, this paper proposes PupuM2D, a musical Masked Modeling Duo (M2D) model that explicitly models music on 2D spectrograms. Instead of relying on 1D tokenization, PupuM2D learns robust representations by predicting the latents of masked spectrogram patches from the unmasked contexts. To optimize such a framework for music, we apply domain-specific modifications to the model architecture, training scheme, and inference paradigm, and conduct comprehensive ablation studies to demonstrate their effectiveness. Evaluations on MARBLEv2 benchmark show that PupuM2D consistently outperforms existing 1D sequence-based baselines across diverse MIR tasks under linear probing; and visualizations of the model's attention maps also confirm that PupuM2D can naturally capture musically meaningful patterns within the 2D time-frequency domain.}

\section{Related Work}
\label{sec:related-work}

\subsection{Visual Self-Supervised Learning}
\label{subsec:rw-vision}

\revision{Vision SSL models have evolved rapidly in recent years.
Masked Autoencoders (MAE)~\cite{mae} first learn powerful visual representations by reconstructing the raw pixels of masked image patches.
However, such pixel-level reconstruction often biases the model toward low-level details rather than high-level semantic representations. 
To address this limitation, JEPAs~\cite{ijepa, vjepa, vjepa2} shift the prediction target from the pixel space to the latent space, encouraging the model to capture robust and semantic visual features without relying on pixel-level losses.
Conversely, contrastive learning, represented by iBOT~\cite{ibot} and the DINO series~\cite{dino, dino2, dino3}, learns representations by maximizing the agreement between differently augmented views of the same image, achieving SOTA performance across various visual downstream tasks.
However, these contrastive methods heavily rely on image-specific augmentations such as cropping, rotation, and color jittering, which do not naturally transfer to music.
While early audio and music contrastive learning methods~\cite{clmr, mule} explore augmentations such as filtering and reverberation, these operations are more effective for isolated tracks than for complete songs, thereby limiting their applicability to full-track music representation learning.}

\subsection{Audio Self-Supervised Learning}
\label{subsec:rw-audio}
\vspace{-4pt}

Inspired by advances in visual SSL, audio SSL has been widely adopted for 2D modeling of Mel-spectrograms.
In particular, AudioMAE~\cite{audiomae, audiomae++} pioneers this by successfully adapting MAE to reconstruct masked spectrogram patches, followed by MSM-MAE~\cite{msmmae}, MaskSpec~\cite{maskspec}, and MAE-AST~\cite{maeast} to further expand the paradigm across various audio tasks.
To overcome the limitations of pixel reconstruction, BEAT~\cite{beat} and EAT~\cite{eat} use acoustic tokenizers to generate discrete prediction targets; \revision{data2vec~\cite{data2vec}, M2D~\cite{m2d, m2dx}, and ATST~\cite{atst} established a latent-space self-distillation paradigm for audio representation learning earlier than the JEPA series, followed by the MATPAC~\cite{audiojepa-design, matpac, matpac++} series to further enhance this paradigm with advanced MoE patch-clustering strategies. Meanwhile, models such as A-JEPA~\cite{ajepa} successfully adapt the JEPA framework to audio, yet their performance still lags behind M2D-style methods on audio domain benchmarks.} More recently, Dasheng~\cite{dasheng} systematically optimizes and scales up the MAE architecture, aiming to build a universal SSL paradigm for general sound.
However, directly adopting audio SSL models to the music domain remains suboptimal.
As their pipelines are designed for audio events, their large patch size will render fine-grained musical events inseparable, and their training scheme will collapse in musical contexts.

\section{Methodology}
\label{sec:method}

\begin{figure*}[t]
    \centering
    \includegraphics[width=\textwidth]{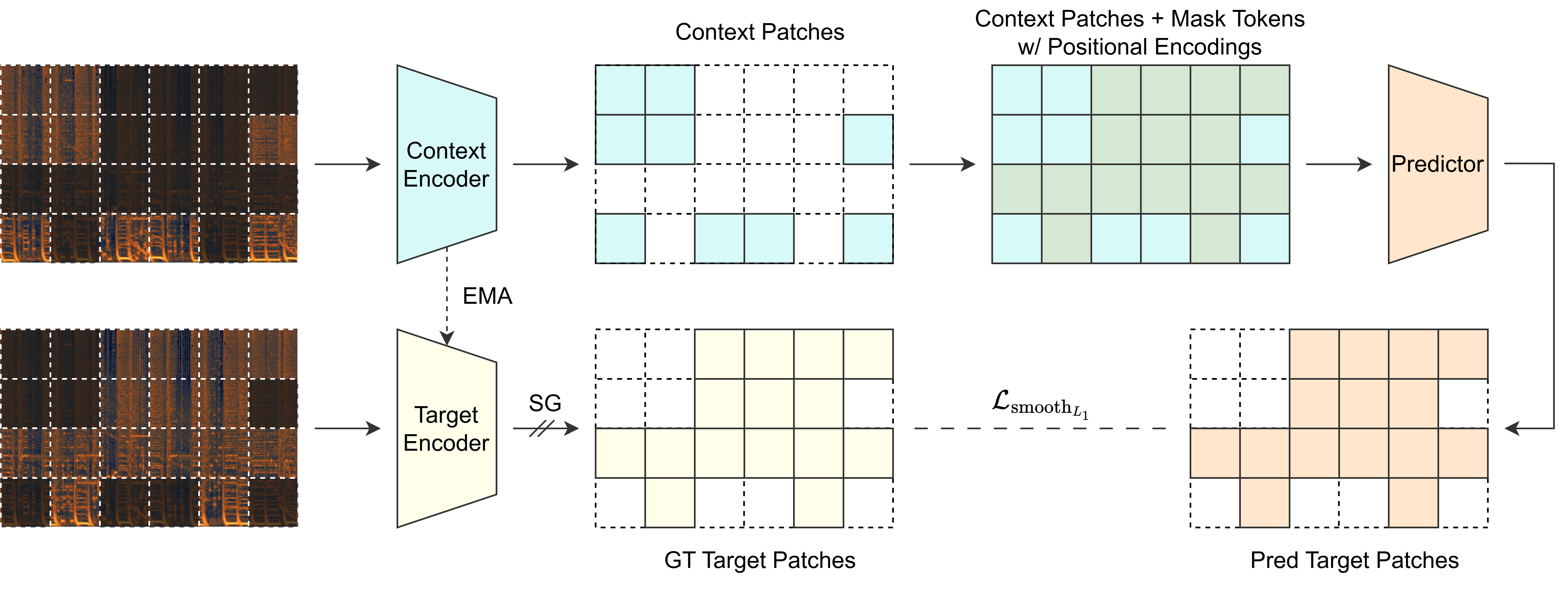}
    \caption{\revision{Overview of the PupuM2D architecture and training scheme. The target encoder encodes the masked spectrogram patches into target latent patches. The context encoder encodes the unmasked spectrogram patches into context latent patches, which are subsequently combined with mask tokens and injected with positional encodings before being passed to the predictor to estimate the target latent patches. The target encoder is updated via an exponential moving average (EMA) of the context encoder, and stop gradient (SG) is applied to prevent representation collapse.}}
    \label{fig:model}
    \vspace{-6pt}
\end{figure*}

\revision{This section details the model architecture, training scheme, and inference paradigm of the proposed PupuM2D framework.} 

\subsection{Model Architecture}

\revision{As illustrated in Figure~\ref{fig:model}, PupuM2D comprises a context encoder, a target encoder, and a predictor, all implemented as standard Vision Transformers (ViTs)~\cite{vit}.} The context encoder extracts representations from the unmasked context patches, and the target encoder extracts representations from the masked target patches. The predictor takes in the context embeddings and mask tokens to predict the target embeddings.

\subsection{Spectrogram Patching}

Following relevant works~\cite{audiomae, audiomae++, msmmae, maskspec, maeast, beat, eat, data2vec, m2d, m2dx, atst, audiojepa-design, matpac, matpac++, ajepa}, we transform music signals into Mel-spectrograms and apply the standard mean-variance normalization to stabilize training. The spectrograms will then be divided into non-overlapping regular-grid patches. Unlike visual SSL models, we use a patch size of 4$\times$16 instead of 16$\times$16 to ensure a suitable frame rate for MIR tasks that require high temporal resolution. These patches are then flattened and embedded by a linear projection. We add rotary position embeddings (RoPE)~\cite{rope} to the embedded patches to accommodate variable-length audio inputs in inference time.

\begin{figure}[t]
    \centering
    \begin{subfigure}[b]{0.32\columnwidth}
         \centering
         \includegraphics[width=\columnwidth]{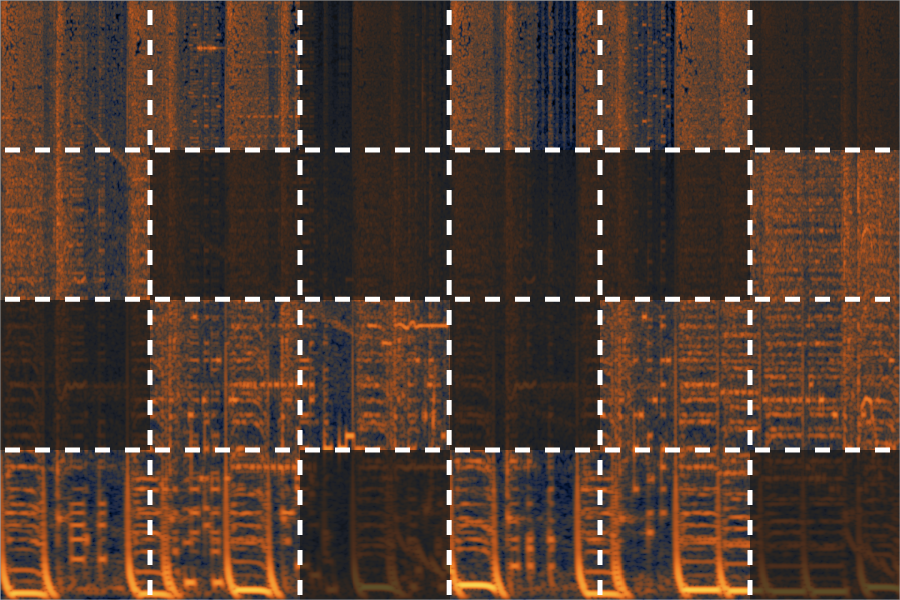}
         \caption{Random}
         \label{fig:random}
    \end{subfigure}
    \hfill
    \begin{subfigure}[b]{0.32\columnwidth}
         \centering
         \includegraphics[width=\columnwidth]{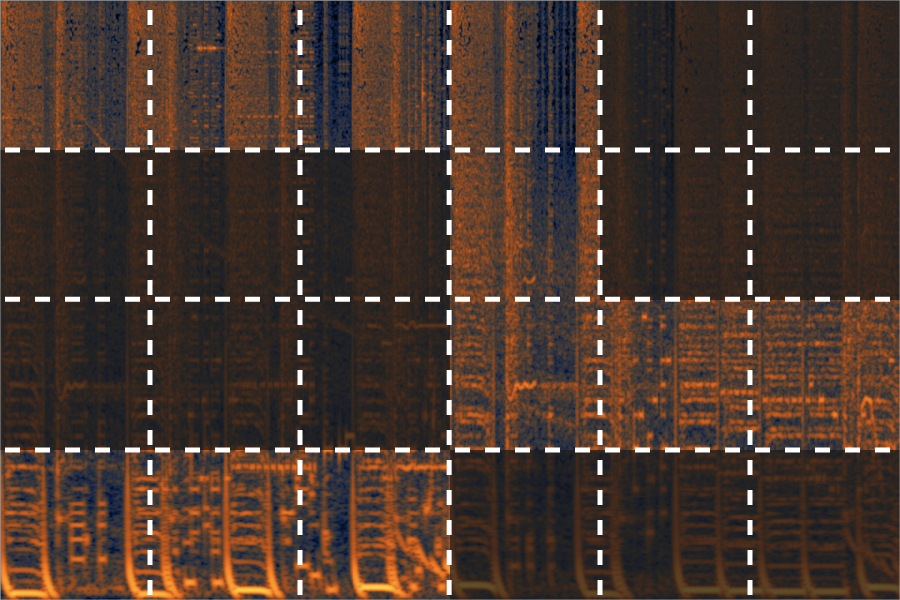}
         \caption{Blockwise}
         \label{fig:block}
    \end{subfigure}
    \hfill
    \begin{subfigure}[b]{0.32\columnwidth}
         \centering
         \includegraphics[width=\columnwidth]{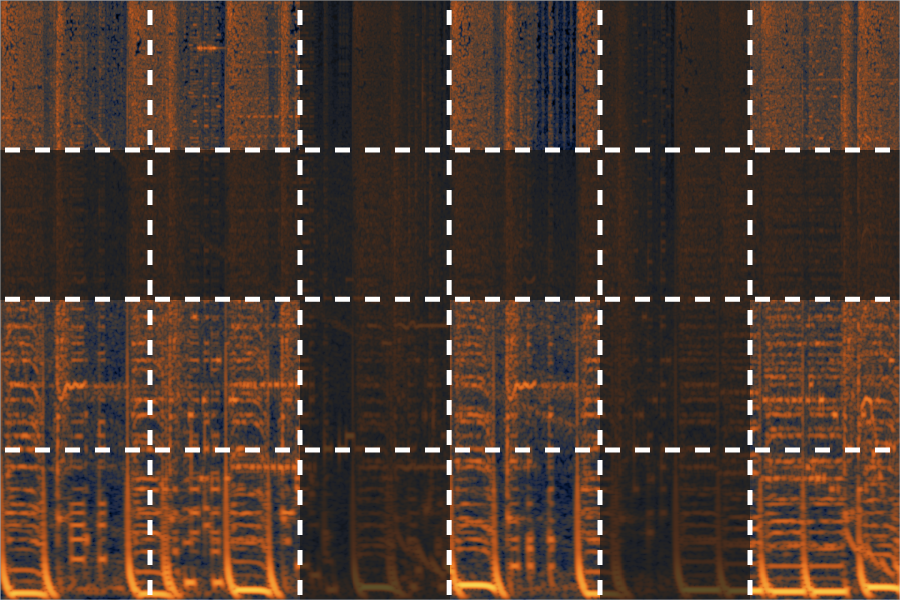}
         \caption{Time-Frequency}
         \label{fig:time-frequency}
    \end{subfigure}
    \caption{Illustration of different masking strategies.}
    \label{fig:masking}
\end{figure}

\subsection{Masking Strategies}

\revision{Different from JEPA implementations~\cite{ijepa, vjepa, vjepa2}, we restrict the target encoder to process only target patches rather than all patches, following M2D~\cite{m2d, m2dx}. This is because music and audio spectrograms exhibit correlated temporal and harmonic structures along the time and frequency axes; exposing the target encoder to the full input patches would allow the model to exploit this information, leading to shortcut learning and representation collapse. For the same reason, we also diverge from existing audio SSL works~\cite{audiomae, audiomae++, msmmae, maskspec, maeast, beat, eat, data2vec, m2d, m2dx, atst, audiojepa-design, matpac, matpac++, ajepa}, where simple random masking is typically sufficient, and additionally use blockwise and time-frequency masking, as shown in Figure~\ref{fig:masking}.}

\subsection{Architectural Optimizations}

We introduce several critical architectural modifications to optimize the ViT backbone's performance on spectrograms. Specifically, we replace the traditional Feed-Forward Network activation with SwiGLU~\cite{swiglu} to enhance model capacity, and apply Query-Key Normalization (QK-Norm) to stabilize the self-attention logits. Conversely, we discard components such as DropPath and LayerScale, as they consistently cause severe training instabilities and lead to representation collapse. Additionally, we replace Batch Normalization with standard Layer Normalization, therefore further stabilizing training.

\subsection{Training Objective}

\revision{PupuM2D is optimized by minimizing the distance between the predictor's outputs and the target encoder's representations across all masked patches. Different from standard M2D implementations~\cite{m2d, m2dx}, we employ a smoothed $L_1$ loss to ensure training stability, which is defined as follows:}
\begin{equation}
    \mathcal{L}_{\text{smooth}L_1}(\hat{z}_{t,f}, z_{t,f}) = 
    \begin{cases} 
        \frac{0.5 (\hat{z}_{t,f} - z_{t,f})^2}{\beta} & \text{if } |\hat{z}_{t,f} - z_{t,f}| < \beta, \\
        |\hat{z}_{t,f} - z_{t,f}| - 0.5\beta & \text{otherwise},
    \end{cases}
\end{equation}
where $\beta=1.0$ is the threshold that controls the loss behaviors; $\hat{z}_{t,f}$ and $z_{t,f}$ are the predictor's output and the target encoder's representation for the masked patch. Note that $z_{t,f}$ is locally mean-variance normalized with mini-batch statistics along the feature dimension, ensuring the loss is not dominated by high-magnitude features and thereby stabilizing training and enhancing the resulting representation quality.

Compared with the standard $L_1$ loss, which is non-differentiable at zero and can cause severe gradient instabilities near the convergence point, the smoothed $L_1$ loss behaves like an $L_2$ penalty for small errors, ensuring smooth gradients at the convergence point; it also transitions to an $L_1$ penalty for large errors, making it significantly less sensitive to outliers and thus preventing gradient explosion during the early stages of training. \revision{As in the standard M2D implementations~\cite{m2d, m2dx}, we apply a stop-gradient operation to the target encoder to prevent shortcut learning and representation collapse, whose parameters are thus updated using an exponential moving average (EMA) of the context encoder's parameters $\bm{\theta}$, as $\bm{\xi} \leftarrow \tau \bm{\xi} + (1 - \tau) \bm{\theta}$, where $\tau$ is the momentum coefficient and $\bm{\xi}$ is the target encoder's parameters, respectively.}

\subsection{Inference Paradigms}
\label{subsec:method-inference}

\begin{figure*}[t]
    \centering
    \includegraphics[width=\textwidth]{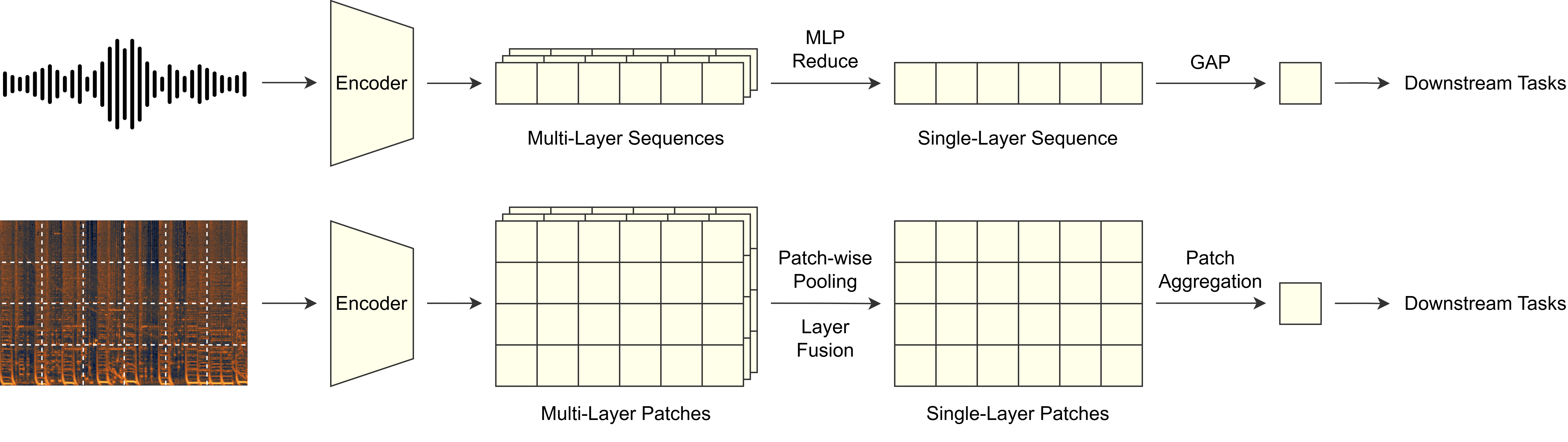}
    \caption{Illustration of different inference paradigms for global tasks. ``GAP'' means global average pooling. The top and bottom panels illustrate the inference paradigms for 1D sequence-based and 2D spectrogram-based models. Note that for fine-grained local tasks, GAP and patch aggregation are omitted, and frame-level features are directly used for downstream tasks.}
    \label{fig:inference}
    \vspace{-8pt}
\end{figure*}

As shown in Figure~\ref{fig:inference}, adapting a 2D spectrogram-based model for downstream MIR tasks is fundamentally different from standard 1D sequence-based baselines. In particular, 1D sequence-based baselines first extract multi-layer 1D hidden representations, which are then applied with an MLP reduction to obtain a single-layer feature, either used directly for fine-grained local tasks or applied with global average pooling (GAP) for global tasks. Such an inference paradigm, however, is incompatible with 2D spectrogram-based models. For the layer fusion strategy, 2D spectrogram-based models produce multi-layer time-frequency patches rather than 1D sequences; flattening and concatenating these patches would result in an exceptionally large feature dimension, making the standard multi-layer MLP reduction incompatible due to parameter explosion. Meanwhile, for the patch aggregation strategy, GAP is essentially suboptimal, as it flattens the spatial layout and results in a complete loss of the essential time-frequency structural information distributed across 2D patches.

\begin{figure}[t]
    \centering
    \begin{subfigure}[b]{0.48\columnwidth}
         \centering
         \includegraphics[width=\columnwidth]{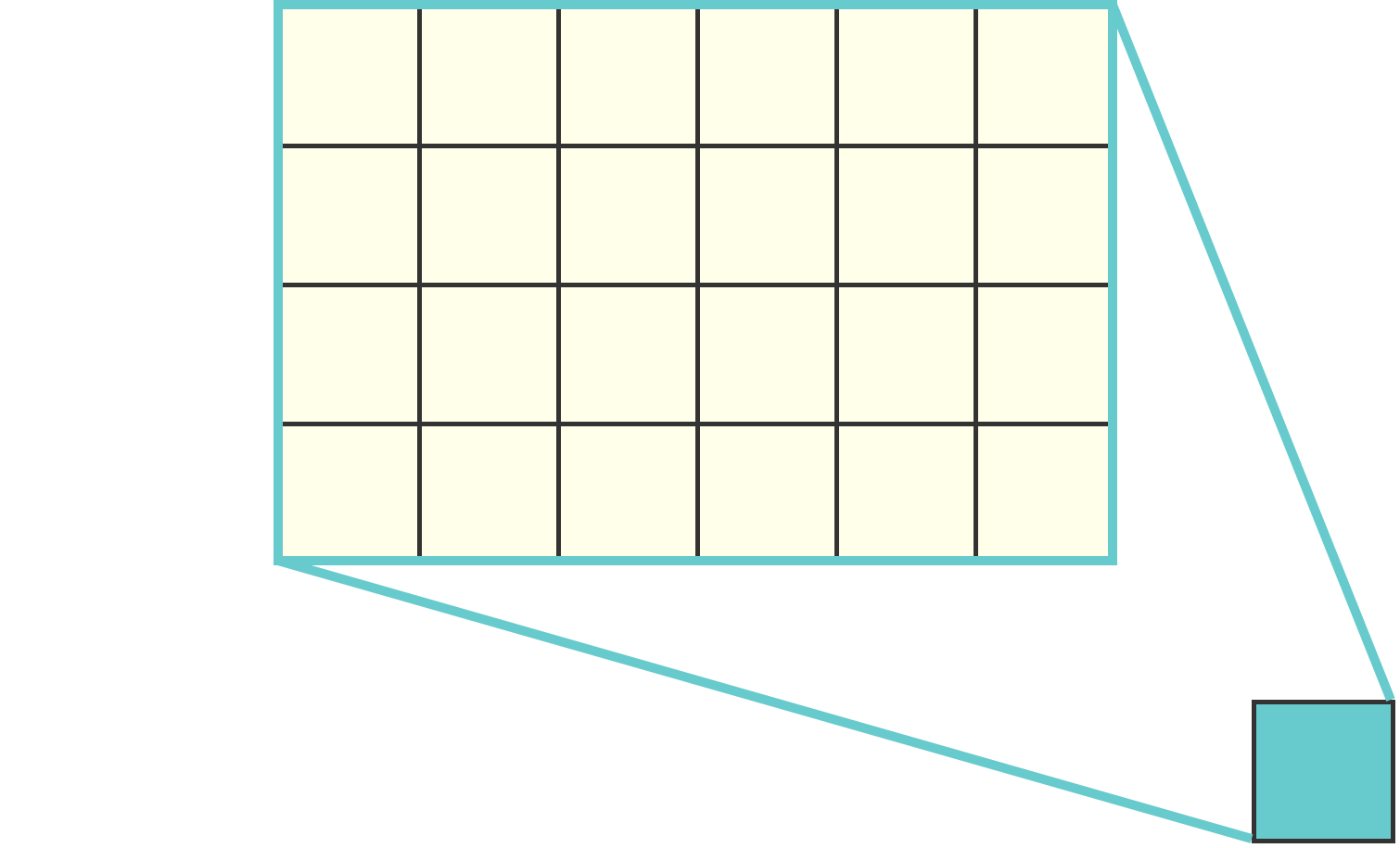}
         \caption{Standard}
         \label{fig:gap-standard}
    \end{subfigure}
    \hfill
    \begin{subfigure}[b]{0.48\columnwidth}
         \centering
         \includegraphics[width=\columnwidth]{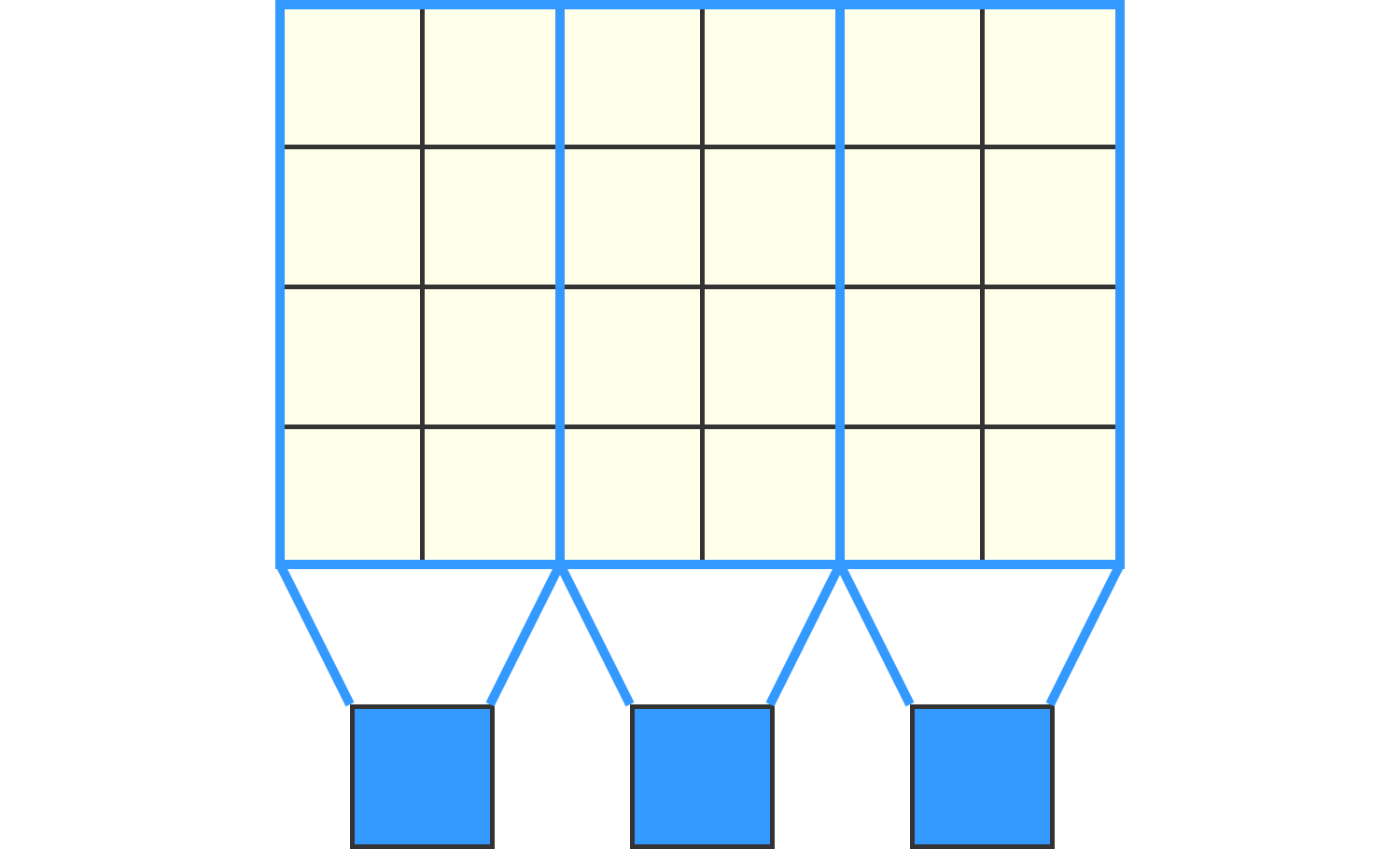}
         \caption{Time-Partitioned}
         \label{fig:gap-time}
    \end{subfigure}
    \\
    \begin{subfigure}[b]{0.48\columnwidth}
         \centering
         \includegraphics[width=\columnwidth]{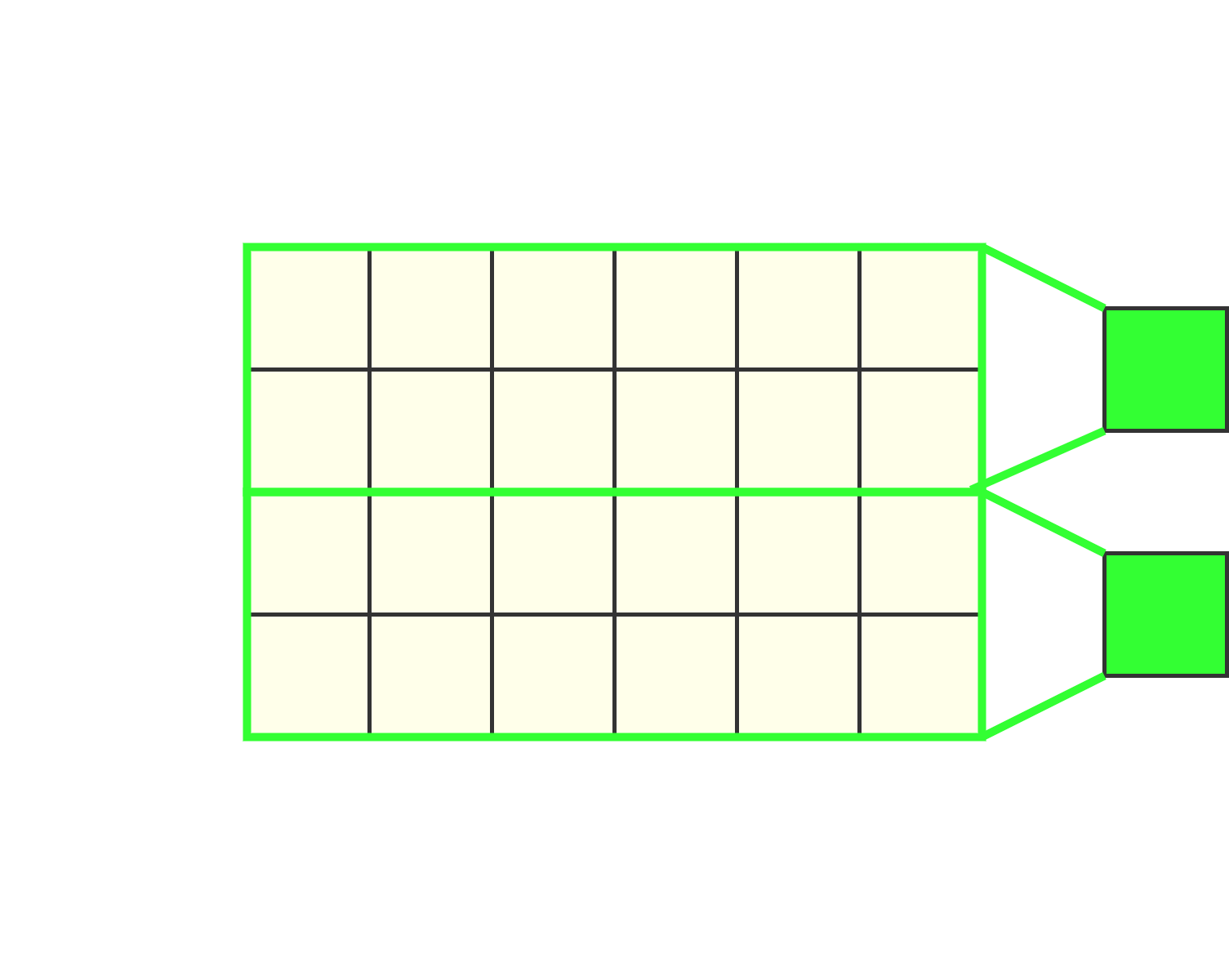}
         \caption{Frequency-Partitioned}
         \label{fig:gap-freq}
    \end{subfigure}
    \hfill
    \begin{subfigure}[b]{0.48\columnwidth}
         \centering
         \includegraphics[width=\columnwidth]{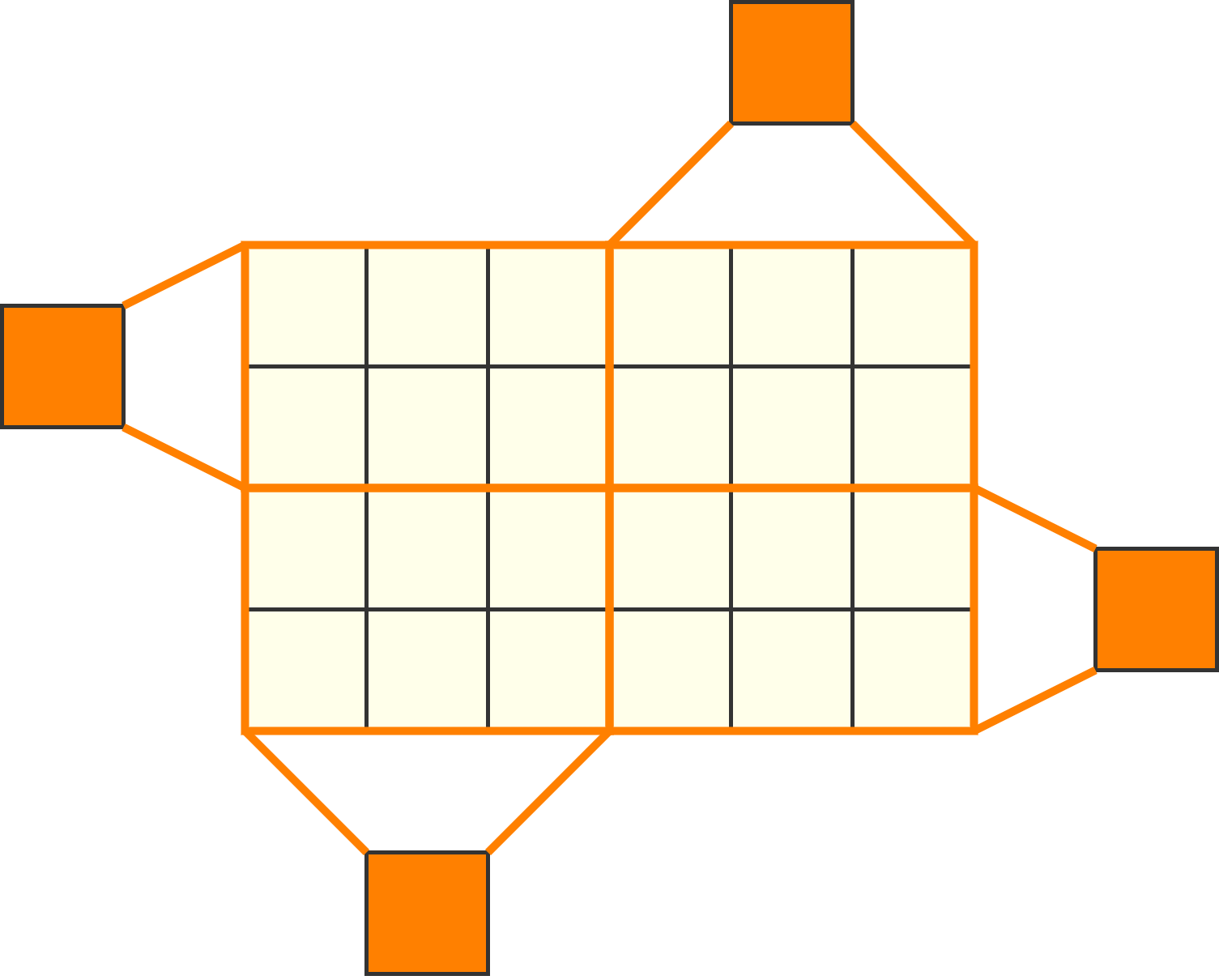}
         \caption{Block-Partitioned}
         \label{fig:gap-block}
    \end{subfigure}
    \caption{Illustration of different patch aggregation strategies.}
    \label{fig:gap}
    \vspace{-14pt}
\end{figure}

To address these issues, we introduce alternative layer fusion and patch aggregation strategies tailored for 2D spectrogram-based models. Let the multi-layer feature extracted from the encoder be $H \in \mathbb{R}^{L \times T \times F \times D}$, where $L$ is the number of layers, $T$ and $F$ represent the number of patches along the time and frequency axes, and $D$ is the hidden dimension. We first apply patch-wise average pooling along the hidden dimension with a pooling factor $p$, yielding $\tilde{H} \in \mathbb{R}^{L \times T \times F \times D'}$, where $D' = D/p$. Then, if no pooling is applied ($p=1$) or $p$ is small, we compute the single layer feature $Z \in \mathbb{R}^{T \times F \times D'}$ with a learnable weighted sum across all the $L$ layers, as: $z_{t, f} = \sum_{l=1}^{L} \bm{w}_l \tilde{h}_{l, t, f}$, where $\bm{w}_l$ are learnable scalar weights normalized via softmax, denoted as \textit{Weighted Sum}. Conversely, if $p$ is large, the compressed hidden dimension will be on par with the standard 1D sequence-based baselines, allowing us to apply the standard multi-layer \textit{MLP Reduce}, which can be written as: $z_{t, f} = \text{MLP}(\text{Concat}(\tilde{h}_{1, t, f}, \tilde{h}_{2, t, f}, \dots, \tilde{h}_{L, t, f}))$.

As illustrated in Figure~\ref{fig:gap}, for global tasks, we additionally introduce a family of time-frequency-structure-aware patch aggregation methods to replace the standard GAP, including \textit{Time-}, \textit{Frequency-}, and \textit{Block-Partitioned Patch Aggregation}. Their detailed formulations are defined in the equations below:

\begin{equation}
    v_{n, m} = \frac{1}{(T/N)(F/M)} \sum_{t=n\frac{T}{N}}^{(n+1)\frac{T}{N}} \sum_{f=m\frac{F}{M}}^{(m+1)\frac{F}{M}} z_{t, f},
\end{equation}
where $n \in \{0, \dots, N-1\}$ and $m \in \{0, \dots, M-1\}$; $N$ and $M$ are the number of chunks along the time/frequency axes. In particular, \textit{Time- ($N>1, M=1$)} and \textit{Frequency-Partitioned ($N=1, M>1$) Patch Aggregation} divide the 2D patches into chunks alongside the time/frequency axes, while \textit{Block-Partitioned ($N>1, M>1$) Patch Aggregation} partitions the 2D patches into localized rectangular blocks to preserve their coarse time-frequency correlations. The resulting global representation $v$ for downstream tasks is obtained by flattening and concatenating the average-pooled $v_{n, m}$ from all blocks:
\begin{equation}
    v = \text{Concat}(v_{1,1}, \dots, v_{N,M}).
\end{equation}

\section{Experiments}

\subsection{Experiment Setup}

\subsubsection{Datasets}

\revision{For PupuM2D training, we use a large-scale, in-house dataset of approximately 100,000 hours of high-quality music, encompassing a highly diverse collection of tracks across various genres, instruments, styles, and keys.}

\subsubsection{Preprocessing}

We process the training dataset to 24\,kHz mono WAV files for simplified model training. For extracting the Mel-spectrograms, we use an FFT size of 1024, a hop size of 240, a window length of 1024, and 128 Mel filters. The extracted spectrograms will be transformed to the log scale, with values smaller than 1e-5 clipped to 0. During training, we randomly crop continuous audio segments of 10.24 seconds, yielding fixed-size 2D spectrogram inputs of shape 1024$\times$128 for the network (both being powers of 2 to leverage optimized PyTorch kernels). Coupled with our asymmetric 4$\times$16 patch size, this configuration results in a frame rate of 25\,Hz.

\begin{table}[t]
\centering
\caption{\revision{ViT configurations of the PupuM2D encoders across different scales. ``Dim'' denotes the hidden dimension, ``MLP Dim'' indicates the inner dimension of the SwiGLU activation, and ``Heads'' denotes the number of attention heads.}}
\vspace{-4pt}
\label{tab:configurations}
\setlength\tabcolsep{5pt}
\begin{tabular}{lccccc}
\toprule
\textbf{Model Scale} & \textbf{\# Params} & \textbf{Layers} & \textbf{Dim} & \textbf{MLP Dim} & \textbf{Heads} \\
\midrule
PupuM2D-Tiny & 5M & 12 & 192 & 768 & 3 \\
PupuM2D-Small & 22M & 12 & 384 & 1536 & 6 \\
PupuM2D-Base & 86M & 12 & 768 & 3072 & 12 \\
PupuM2D-Large & 307M & 24 & 1024 & 4096 & 16 \\
PupuM2D-Huge & 632M & 32 & 1280 & 5120 & 16 \\
\bottomrule
\end{tabular}
\vspace{-26pt}
\end{table}

\subsubsection{Configurations}

\revision{We implement PupuM2D across five parameter scales: Tiny, Small, Base, Large, and Huge. The detailed ViT configurations of the encoder across all parameter scales are illustrated in Table~\ref{tab:configurations}.} 
Following relevant audio SSL works~\cite{audiomae, audiomae++, msmmae, maskspec, maeast, beat, eat, data2vec, m2d, m2dx, atst, audiojepa-design, matpac, matpac++, ajepa}, the predictor is fixed across all model scales and is implemented as a lightweight 8-layer ViT with a hidden dimension of 512, an MLP dimension of 2048, and 16 attention heads. Note that the parameter counts reported in our experiments include only the target encoder, which serves as the standalone feature extractor for downstream tasks. 

As for masking strategies, we set the overall target masking ratio to $\rho = 0.8$ following previous works~\cite{audiomae, audiomae++, msmmae, maskspec, maeast, beat, eat, data2vec, m2d, m2dx, atst, audiojepa-design, matpac, matpac++, ajepa} and dynamically apply one of three masking strategies for each input, with selection probabilities $P_{\text{rand}} = 0.4$, $P_{\text{block}} = 0.3$, and $P_{\text{time-freq}} = 0.3$. The detailed implementations are:

\begin{itemize}
    \item \textbf{Random Masking:} Individual patches are uniformly sampled and removed across the entire time-frequency grid until the target masking ratio $\rho$ is reached.
    
    \item \textbf{Blockwise Masking:} We iteratively sample rectangular blocks to mask. For each block, the area scale $s$ relative to the full grid is sampled uniformly with $s \sim \mathcal{U}(0.01, 0.05)$. To ensure the model learns both long-temporal and wide-spectral patterns, the block aspect ratio $r$ is sampled from a bimodal distribution, chosen equally from either a vertical range $r \sim \mathcal{U}(2.0, 5.0)$ or a horizontal range $r \sim \mathcal{U}(0.2, 0.5)$. To precisely achieve the target ratio $\rho$ without over-masking, we halt block generation once adding a new block would exceed the limit, and fulfill the residual masking requirement by uniformly sampling individual patches, as in the random masking.
    
    \item \textbf{Time-Frequency Masking:} We iteratively mask chunk-wise spans along the time/frequency axes. To compensate for the large area covered by frequency masking, the probability of selecting an axis is proportional to its span size, as $P_{\text{time}} = \frac{T}{T+F}$ and $P_{\text{freq}} = \frac{F}{T+F}$. We then remove temporal and spectral chunks up to a maximum width of $w_{\text{time}} = 40$ and $w_{\text{freq}} = 1$ patches. Similar to blockwise masking, we halt chunk generation once the limit is reached and apply random masking afterward.
\end{itemize}

\begin{table}[t]
\centering
\caption{\revision{Layer fusion selection for downstream tasks of the PupuM2D encoders at different model scales.}}
\vspace{-4pt}
\label{tab:selected-layers}
\begin{tabular}{lcc}
\toprule
\textbf{Model Scale} & \textbf{Layers} & \textbf{Selected Layer Indices} \\
\midrule
PupuM2D-Tiny  & 12 & $\{9, 10, \dots, 12\}$ \\
PupuM2D-Small & 12 & $\{9, 10, \dots, 12\}$ \\
PupuM2D-Base  & 12 & $\{9, 10, \dots, 12\}$ \\
PupuM2D-Large & 24 & $\{12, 13, \dots, 24\}$ \\
PupuM2D-Huge  & 32 & $\{16, 17, \dots, 32\}$ \\
\bottomrule
\end{tabular}
\vspace{-16pt}
\end{table}

\subsubsection{Training}

\revision{We train all PupuM2D models for a total of 500,000 steps with a global batch size of 2048 on 32 NVIDIA B200 GPUs.} We adapted BF16 mixed precision to maximize computational efficiency. The networks are optimized using the AdamW optimizer. To stabilize training, all gradients are clipped to a maximum norm of 1.0, and a constant weight decay of 0.05 is applied to all weights (explicitly excluding biases, normalization layers, learnable tokens, and positional embeddings). The learning rate uses a cosine decay schedule, warming up linearly over the first $T_{w} = 25,000$ steps to a peak of $1 \times 10^{-3}$, and subsequently decaying to $1 \times 10^{-6}$. The target encoder is updated using EMA, with the momentum coefficient increasing linearly from 0.99995 to 0.99999. 

To ease the optimization difficulty in the early training phase, we introduce a curriculum-based masking scheduling mechanism. Let $\text{P}(t) = (P_{\text{rand}}, P_{\text{block}}, P_{\text{time-freq}})$ denote the selection probabilities for the random, blockwise, and time-frequency masking. Here, we define the initial purely random distribution as $\text{P}_{\text{init}} = (1.0, 0, 0)$, and target mixed distribution as $\text{P}_{\text{target}} = (0.4, 0.3, 0.3)$. The dynamic masking distribution transitions over a specified $T_{m} = 100,000$ steps:
\begin{equation}
    \text{P}(t) = 
    \begin{cases}
        \text{P}_{\text{init}} & t \le T_{w}, \\
        \text{P}_{\text{init}} + (\text{P}_{\text{target}} - \text{P}_{\text{init}}) \frac{t - T_{w}}{T_{m} - T_{w}} & T_{w} < t \le T_{m}, \\
        \text{P}_{\text{target}} & t > T_{m}.
    \end{cases}
\end{equation}

\begin{table*}[t]
\begin{center}
\caption{\revision{Experimental results of PupuM2D and baseline models in the MARBLEv2 benchmark (1/2). ``*'' denotes the results are obtained by evaluating official checkpoints provided by their papers within our evaluation environment, and ``$\ddagger$'' denotes the results are obtained from audio-domain baseline models that we reproduced and pre-trained on our datasets under identical settings. Note that the local task evaluation results of some audio and universal SSL models are omitted, as their output feature frame rates are insufficient. The best and second-best results for each column are \textbf{bold} and \underline{underlined}, respectively.}}
\label{tab:results-marble-1}
\setlength\tabcolsep{4pt}
\begin{tabular}{clccccccccc}
\toprule
& \textbf{Dataset} & \multicolumn{2}{c}{\textbf{EMO}} & \textbf{GS} & \multicolumn{2}{c}{\textbf{GTZAN}} & \multicolumn{2}{c}{\textbf{HookTheory}} & \multicolumn{2}{c}{\textbf{MTT}} \\ 
& \textbf{Type} & \multicolumn{2}{c}{\textit{Global}} & \textit{Local} & \textit{Global} & \textit{Local} & \textit{Local} & \textit{Local} & \multicolumn{2}{c}{\textit{Global}} \\
& \textbf{Task} & \multicolumn{2}{c}{Emotion} & Key & Genre & Rhythm & Key & Structure & \multicolumn{2}{c}{Tagging} \\
\midrule
\midrule
\textbf{\# Param} & \textbf{Model} & R2$^{\text{V}}$ ($\uparrow$) & R2$^{\text{A}}$ ($\uparrow$) & Acc$^{\text{Refined}}$ ($\uparrow$) & Acc ($\uparrow$) & F1$^{\text{Beat}}$ ($\uparrow$) & Acc$^{\text{Refined}}$ ($\uparrow$) & Acc ($\uparrow$) & ROC ($\uparrow$) & AP ($\uparrow$) \\
\midrule
95M & MERT-Base$^*$~\cite{mert} & 54.1$^*$ & 75.3$^*$ & 62.5$^*$ & 78.3$^*$ & 87.9$^*$ & \textbf{73.4}$^*$ & 55.0$^*$ & 90.3$^*$ & 37.5$^*$ \\
330M & MERT-Large$^*$~\cite{mert} & 56.7$^*$ & 76.1$^*$ & 64.1$^*$ & 77.6$^*$ & 86.8$^*$ & 70.4$^*$ & 54.8$^*$ & 90.6$^*$ & 37.9$^*$ \\
310M & MuQ$^*$~\cite{muq} & 58.3$^*$ & 76.4$^*$ & 63.2$^*$ & 83.8$^*$ & 90.1$^*$ & 72.8$^*$ & 57.0$^*$ & 90.5$^*$ & 38.5$^*$ \\
330M & MusicFM$^*$~\cite{musicfm} & 57.2$^*$ & 74.4$^*$ & 63.0$^*$ & 84.1$^*$ & 90.2$^*$ & 71.8$^*$ & \textbf{59.8}$^*$ & 90.9$^*$ & 38.3$^*$ \\
\midrule
9M & AudioMAE++-Tiny$^*$~\cite{audiomae, audiomae++} & 52.4$^*$ & 71.2$^*$ & / & 73.4$^*$ & / & / & / & 91.2$^*$ & 40.2$^*$ \\
142M & AudioMAE++-Base$^*$~\cite{audiomae, audiomae++} & 57.0$^*$ & 72.4$^*$ & / & 80.7$^*$ & / & / & / & 91.6$^*$ & 40.4$^*$ \\
504M & AudioMAE++-Large$^*$~\cite{audiomae, audiomae++} & 54.6$^*$ & 74.1$^*$ & / & 84.5$^*$ & / & / & / & \underline{91.9}$^*$ & \underline{41.1}$^*$ \\
86M & M2D$^*$~\cite{m2d} & 56.4$^*$ & \underline{77.6}$^*$ & / & 81.7$^*$ & / & / & / & 91.8$^*$ & 40.9$^*$ \\
90M & BEATs$^*$~\cite{beat} & 58.1$^*$ & 73.1$^*$ & / & 80.0$^*$ & / & / & / & 91.4$^*$ & 39.6$^*$ \\
\midrule
86M & MATPAC++-$^*$~\cite{audiojepa-design, matpac, matpac++} & 59.3$^*$ & 77.1$^*$ & / & \textbf{87.2}$^*$ & / & / & / & \textbf{92.0}$^*$ & \textbf{41.7}$^*$ \\
86M & M2D-X$^*$~\cite{m2dx} & 57.6$^*$ & 75.6$^*$ & / & 86.6$^*$ & / & / & / & \underline{91.9}$^*$ & 40.9$^*$ \\
86M & Dasheng-Base$^*$~\cite{dasheng} & 59.6$^*$ & 75.7$^*$ & 54.5$^*$ & 76.9$^*$ & 87.2$^*$ & 63.1$^*$ & 57.3$^*$ & 91.5$^*$ & 40.6$^*$ \\
600M & Dasheng-0.6B$^*$~\cite{dasheng} & 59.0$^*$ & 76.7$^*$ & 55.4$^*$ & 81.4$^*$ & 88.2$^*$ & 66.9$^*$ & 57.9$^*$ & 91.6$^*$ & 40.3$^*$ \\
1.2B & Dasheng-1.2B$^*$~\cite{dasheng} & 57.4$^*$ & 75.0$^*$ & 58.0$^*$ & 81.4$^*$ & 87.7$^*$ & 67.0$^*$ & 57.5$^*$ & 91.5$^*$ & 40.4$^*$ \\
\midrule
307M & AudioMAE++-Large$^\ddagger$~\cite{audiomae, audiomae++} & 59.0$^\ddagger$ & 75.7$^\ddagger$ & 61.7$^\ddagger$ & 80.3$^\ddagger$ & 90.0$^\ddagger$ & 72.2$^\ddagger$ & 57.3$^\ddagger$ & 91.2$^\ddagger$ & 39.5$^\ddagger$ \\
307M & MATPAC++-Large$^\ddagger$~\cite{audiojepa-design, matpac, matpac++} & 57.8$^\ddagger$ & 74.7$^\ddagger$ & 63.7$^\ddagger$ & 81.4$^\ddagger$ & 90.1$^\ddagger$ & 70.6$^\ddagger$ & 56.6$^\ddagger$ & 90.6$^\ddagger$ & 38.2$^\ddagger$ \\
307M & M2D-Large$^\ddagger$~\cite{m2d, m2dx} & 57.4$^\ddagger$ & 74.8$^\ddagger$ & \underline{65.0}$^\ddagger$ & 83.8$^\ddagger$ & 90.0$^\ddagger$ & 71.5$^\ddagger$ & 57.8$^\ddagger$ & 91.0$^\ddagger$ & 39.2$^\ddagger$ \\
\midrule
5M & PupuM2D-Tiny & 55.3 & 74.8 & 64.2 & 76.2 & 88.4 & 70.3 & 55.7 & 90.8 & 38.8 \\
22M & PupuM2D-Small & 61.1 & 74.9 & 64.8 & 78.6 & 89.0 & 70.7 & 57.6 & 91.4 & 40.4 \\
86M & PupuM2D-Base & 58.8 & 76.3 & 60.5 & 77.9 & \textbf{91.0} & 71.0 & 57.5 & 91.3 & 39.7 \\
307M & PupuM2D-Large & \textbf{62.5} & 76.8 & \textbf{66.1} & \underline{86.9} & \textbf{91.0} & \underline{72.9} & 57.6 & 91.7 & 40.8 \\
632M & PupuM2D-Huge & \underline{62.0} & \textbf{78.5} & 64.8 & 85.9 & \underline{90.5} & 72.2 & \underline{58.7} & 91.3 & 39.7 \\
\bottomrule
\end{tabular}
\end{center}
\vspace{-8pt}
\end{table*}

\begin{table*}[t]
\begin{center}
\caption{\revision{Ablation study of different training strategies and model architectures on PupuM2D-Large in MARBLEv2 benchmark (1/2). The best and second-best results for each column are \textbf{bold} and \underline{underlined}, respectively.}}
\label{tab:results-ablation-architecture-1}
\begin{tabular}{lccccccccc}

\toprule
\textbf{Dataset} & \multicolumn{2}{c}{\textbf{EMO}} & \textbf{GS} & \multicolumn{2}{c}{\textbf{GTZAN}} & \multicolumn{2}{c}{\textbf{HookTheory}} & \multicolumn{2}{c}{\textbf{MTT}} \\ 
\textbf{Type} & \multicolumn{2}{c}{\textit{Global}} & \textit{Local} & \textit{Global} & \textit{Local} & \textit{Local} & \textit{Local} & \multicolumn{2}{c}{\textit{Global}} \\
\textbf{Task} & \multicolumn{2}{c}{Emotion} & Key & Genre & Rhythm & Key & Structure & \multicolumn{2}{c}{Tagging} \\
\midrule
\midrule
\textbf{Model} & R2$^{\text{V}}$ ($\uparrow$) & R2$^{\text{A}}$ ($\uparrow$) & Acc$^{\text{Refined}}$ ($\uparrow$) & Acc ($\uparrow$) & F1$^{\text{Beat}}$ ($\uparrow$) & Acc$^{\text{Refined}}$ ($\uparrow$) & Acc ($\uparrow$) & ROC ($\uparrow$) & AP ($\uparrow$) \\
\midrule
PupuM2D-Large & \textbf{62.5} & 76.8 & \textbf{66.1} & \textbf{86.9} & \textbf{91.0} & \textbf{72.9} & \textbf{57.6} & \textbf{91.7} & \textbf{40.8} \\
\quad w/o SwiGLU & 58.1 & 76.3 & 62.5 & \underline{83.8} & 90.2 & 69.6 & \underline{57.3} & 90.9 & 39.1 \\
\quad w/o QKNorm & \underline{60.8} & \underline{77.3} & 64.1 & \underline{83.8} & \underline{90.8} & 70.0 & 56.5 & \underline{91.3} & \underline{39.3} \\
\quad w/o Mixing Masking Strategy & \underline{60.8} & \textbf{79.1} & \underline{64.4} & 82.8 & \underline{90.8} & \underline{72.8} & \textbf{57.6} & 91.0 & 39.2 \\
\quad w/o Smoothed $L_1$ Loss & \multicolumn{9}{c}{\multirow{5}{*}{\textit{Training Collapsed}}} \\
\quad w/ Full-Patch Target Encoder & \multicolumn{9}{c}{} \\
\quad w/ DropPath & \multicolumn{9}{c}{} \\
\quad w/ LayerScale & \multicolumn{9}{c}{} \\
\quad w/ Batch Normalization & \multicolumn{9}{c}{} \\
\bottomrule

\end{tabular}
\end{center}
\vspace{-15pt}
\end{table*}
    
\subsubsection{Baselines}

\revision{We compare PupuM2D against music SSL baselines, including MERT~\cite{mert}, MusicFM~\cite{musicfm}, and MuQ~\cite{muq}, general audio SSL baselines, including AudioMAE++~\cite{audiomae, audiomae++}, M2D~\cite{m2d}, and BEATs~\cite{beat}, and universal sound SSL baselines, including MATPAC++~\cite{audiojepa-design, matpac, matpac++}, M2D-X~\cite{m2dx}, and Dasheng~\cite{dasheng}. We first report the evaluation results using their official checkpoints and implementations within our evaluation environment. We further reproduced several representative audio and universal SSL models from scratch on our in-house dataset (AudioMAE++-Large~\cite{audiomae++}, MATPAC++-Large~\cite{matpac++}, and M2D-Large~\cite{m2d}). To ensure a fair comparison, we use the exact same optimal ViT model architecture and inference paradigm developed for PupuM2D (i.e., layer fusion and patch aggregation strategies) for both training and evaluation.}

\subsection{Downstream Tasks}
\vspace{-11pt}

\revision{We use MARBLEv2~\cite{marble} benchmark to evaluate PupuM2D and baseline models on MIR tasks using linear probing. \textbf{Note that the scores derived from MARBLEv2 are not strictly comparable to those from MARBLEv1 due to differences in the benchmark implementation}. Following~\cite{layer-selection}, we select intermediate layers that capture rich musical semantics and exhibit strong linear separability for evaluation, as illustrated in Table~\ref{tab:selected-layers}. All evaluations are conducted on NVIDIA B200 GPUs. 
In particular, we broadly categorize these tasks into two types: \textit{global tasks} and \textit{local tasks}. \textit{global tasks} aim to predict track-level labels such as emotion, genre, and tags, while fine-grained \textit{local tasks} aim to predict frame-level labels such as musical key and beat types. The detailed task descriptions are:}

\begin{table*}[t]
\begin{center}
\caption{\revision{Experimental results of PupuM2D and baseline models in the MARBLEv2 benchmark (2/2). ``*'' denotes the results are obtained by evaluating official checkpoints provided by their papers within our evaluation environment, and ``$\ddagger$'' denotes the results are obtained from audio-domain baseline models that we reproduced and pre-trained on our datasets under identical settings. Note that the local task evaluation results of some audio and universal SSL models are omitted, as their output feature frame rates are insufficient. The best and second-best results for each column are \textbf{bold} and \underline{underlined}, respectively.}}
\label{tab:results-marble-2}
\begin{tabular}{clcccccccc}

\toprule
& \textbf{Dataset} & \multicolumn{2}{c}{\textbf{MTG}} & \multicolumn{2}{c}{\textbf{MTG}} & \multicolumn{2}{c}{\textbf{MTG}} & \multicolumn{2}{c}{\textbf{MTG}} \\ 
& \textbf{Type} & \multicolumn{2}{c}{\textit{Global}} & \multicolumn{2}{c}{\textit{Global}} & \multicolumn{2}{c}{\textit{Global}} & \multicolumn{2}{c}{\textit{Global}} \\
& \textbf{Task} & \multicolumn{2}{c}{Instrument} & \multicolumn{2}{c}{MoodTheme} & \multicolumn{2}{c}{Genre} & \multicolumn{2}{c}{Top50} \\
\midrule
\midrule
\textbf{\# Param} & \textbf{Model} & ROC ($\uparrow$) & AP ($\uparrow$) & ROC ($\uparrow$) & AP ($\uparrow$) & ROC ($\uparrow$) & AP ($\uparrow$) & ROC ($\uparrow$) & AP ($\uparrow$) \\
\midrule
95M & MERT-Base$^*$~\cite{mert} & 77.1$^*$ & 19.7$^*$ & 75.8$^*$ & 13.5$^*$ & 86.2$^*$ & 18.4$^*$ & 81.8$^*$ & 28.8$^*$ \\
330M & MERT-Large$^*$~\cite{mert} & 75.5$^*$ & 18.8$^*$ & 75.3$^*$ & 13.5$^*$ & 86.1$^*$ & 18.0$^*$ & 82.6$^*$ & 29.1$^*$ \\
310M & MuQ$^*$~\cite{muq} & 74.8$^*$ & 19.1$^*$ & 73.7$^*$ & 13.2$^*$ & 85.4$^*$ & 19.1$^*$ & 83.0$^*$ & 30.2$^*$ \\
330M & MusicFM$^*$~\cite{musicfm} & 74.6$^*$ & 18.5$^*$ & 74.9$^*$ & 14.1$^*$ & 85.3$^*$ & 19.4$^*$ & 81.9$^*$ & 29.7$^*$ \\
\midrule
9M & AudioMAE++-Tiny$^*$~\cite{audiomae, audiomae++} & 76.8$^*$ & 19.4$^*$ & 76.2$^*$ & 14.5$^*$ & 85.8$^*$ & 17.3$^*$ & 82.3$^*$ & 28.0$^*$ \\
142M & AudioMAE++-Base$^*$~\cite{audiomae, audiomae++} & 77.0$^*$ & 20.6$^*$ & 76.6$^*$ & 15.2$^*$ & \textbf{86.5}$^*$ & 19.7$^*$ & 82.8$^*$ & 30.3$^*$ \\
504M & AudioMAE++-Large$^*$~\cite{audiomae, audiomae++} & 76.7$^*$ & 20.6$^*$ & \underline{77.3}$^*$ & 15.9$^*$ & \underline{86.4}$^*$ & 20.1$^*$ & 82.9$^*$ & \underline{30.7}$^*$ \\
86M & M2D$^*$~\cite{m2d} & 77.6$^*$ & \underline{21.1}$^*$ & 76.8$^*$ & 15.9$^*$ & \textbf{86.5}$^*$ & \underline{20.2}$^*$ & \underline{83.1}$^*$ & 30.2$^*$ \\
90M & BEATs$^*$~\cite{beat} & 77.3$^*$ & 20.5$^*$ & 76.0$^*$ & 14.9$^*$ & 86.0$^*$ & 18.5$^*$ & 82.4$^*$ & 29.0$^*$ \\
\midrule
86M & MATPAC++$^*$~\cite{audiojepa-design, matpac, matpac++} & \underline{78.2}$^*$ & \underline{21.1}$^*$ & 77.2$^*$ & \underline{16.1}$^*$ & 86.3$^*$ & \textbf{20.7}$^*$ & \underline{83.1}$^*$ & \underline{30.7}$^*$ \\
86M & M2D-X$^*$~\cite{m2dx} & 77.2$^*$ & 20.9$^*$ & \textbf{77.5}$^*$ & \textbf{16.4}$^*$ & \underline{86.4}$^*$ & 20.1$^*$ & 82.9$^*$ & 30.5$^*$ \\
86M & Dasheng-Base$^*$~\cite{dasheng} & 76.5$^*$ & 19.5$^*$ & 76.9$^*$ & 15.4$^*$ & 86.2$^*$ & 19.0$^*$ & 82.4$^*$ & 29.6$^*$ \\
600M & Dasheng-0.6B$^*$~\cite{dasheng} & 75.9$^*$ & 19.7$^*$ & 77.2$^*$ & 15.9$^*$ & 86.1$^*$ & 19.4$^*$ & 82.7$^*$ & 30.0$^*$ \\
1.2B & Dasheng-1.2B$^*$~\cite{dasheng} & 75.0$^*$ & 19.0$^*$ & 76.1$^*$ & 15.5$^*$ & 85.5$^*$ & 18.8$^*$ & 82.4$^*$ & 29.6$^*$ \\
\midrule
307M & AudioMAE++-Large$^\ddagger$~\cite{audiomae, audiomae++} & 77.1$^\ddagger$ & 19.9$^\ddagger$ & 75.6$^\ddagger$ & 14.0$^\ddagger$ & 86.3$^\ddagger$ & 18.9$^\ddagger$ & \underline{83.1}$^\ddagger$ & \textbf{31.1}$^\ddagger$ \\
307M & MATPAC++-Large$^\ddagger$~\cite{audiojepa-design, matpac, matpac++} & 77.2$^\ddagger$ & 19.7$^\ddagger$ & 75.1$^\ddagger$ & 14.1$^\ddagger$ & 85.7$^\ddagger$ & 19.6$^\ddagger$ & 82.5$^\ddagger$ & 30.2$^\ddagger$ \\
307M & M2D-Large$^\ddagger$~\cite{m2d, m2dx} & 76.6$^\ddagger$ & 19.3$^\ddagger$ & 74.6$^\ddagger$ & 14.3$^\ddagger$ & 85.5$^\ddagger$ & 19.2$^\ddagger$ & 82.5$^\ddagger$ & 29.6$^\ddagger$ \\
\midrule
5M & PupuM2D-Tiny & 77.0 & 19.0 & 74.9 & 13.8 & 85.1 & 17.6 & 82.4 & 28.8 \\
22M & PupuM2D-Small & 77.0 & 19.8 & 75.6 & 14.3 & 85.2 & 18.6 & \textbf{83.2} & 30.1 \\
86M & PupuM2D-Base & 76.2 & 19.6 & 75.7 & 14.2 & 85.7 & 19.4 & 82.7 & 29.7 \\
307M & PupuM2D-Large & \textbf{78.4} & \textbf{21.2} & 76.2 & 15.3 & 86.1 & 20.1 & 82.8 & 30.5 \\
632M & PupuM2D-Huge & 77.6 & 20.5 & 75.9 & 14.7 & 85.9 & 20.1 & \underline{83.1} & \underline{30.7} \\
\bottomrule

\end{tabular}
\end{center}
\vspace{-6pt}
\end{table*}

\begin{table*}[t]
\begin{center}
\caption{\revision{Ablation study of different training strategies and model architectures on PupuM2D-Large in MARBLEv2 benchmark (2/2). The best and second-best results for each column are \textbf{bold} and \underline{underlined}, respectively.}}
\label{tab:results-ablation-architecture-2}
\begin{tabular}{lcccccccc}

\toprule
\textbf{Dataset} & \multicolumn{2}{c}{\textbf{MTG}} & \multicolumn{2}{c}{\textbf{MTG}} & \multicolumn{2}{c}{\textbf{MTG}} & \multicolumn{2}{c}{\textbf{MTG}} \\ 
\textbf{Type} & \multicolumn{2}{c}{\textit{Global}} & \multicolumn{2}{c}{\textit{Global}} & \multicolumn{2}{c}{\textit{Global}} & \multicolumn{2}{c}{\textit{Global}} \\
\textbf{Task} & \multicolumn{2}{c}{Instrument} & \multicolumn{2}{c}{MoodTheme} & \multicolumn{2}{c}{Genre} & \multicolumn{2}{c}{Top50} \\
\midrule
\midrule
\textbf{Model} & ROC ($\uparrow$) & AP ($\uparrow$) & ROC ($\uparrow$) & AP ($\uparrow$) & ROC ($\uparrow$) & AP ($\uparrow$) & ROC ($\uparrow$) & AP ($\uparrow$) \\
\midrule
PupuM2D-Large & \textbf{78.4} & \textbf{21.2} & \textbf{76.2} & \textbf{15.3} & \textbf{86.1} & \textbf{20.1} & \textbf{82.8} & \textbf{30.5} \\
\quad w/o SwiGLU & 76.6 & 19.3 & 74.6 & 14.3 & 85.5 & 19.2 & \underline{82.5} & 29.6 \\
\quad w/o QKNorm & \underline{77.7} & 19.5 & 75.2 & \underline{14.4} & 85.2 & 18.6 & \underline{82.5} & 29.4 \\
\quad w/o Mixing Masking Strategy & 77.1 & \underline{20.0} & \underline{75.4} & 14.3 & \underline{85.6} & \underline{19.8} & 82.0 & \underline{29.7} \\
\quad w/o Smoothed $L_1$ Loss & \multicolumn{8}{c}{\multirow{5}{*}{\textit{Training Collapsed}}} \\
\quad w/ Full-Patch Target Encoder & \multicolumn{8}{c}{} \\
\quad w/ DropPath & \multicolumn{8}{c}{} \\
\quad w/ LayerScale & \multicolumn{8}{c}{} \\
\quad w/ Batch Normalization & \multicolumn{8}{c}{} \\
\bottomrule

\end{tabular}
\end{center}
\vspace{-14pt}
\end{table*}

\begin{table*}[p]
\begin{center}
\caption{\revision{Ablation study of different layer fusion and patch aggregation strategies with various pooling factors on PupuM2D-Large in MARBLEv2's global tasks (1/2). ``Standard'', ``Time-$x$'', ``Freq-$y$'', and ``Block-$x$$\times$$y$'' mean Standard, Time-Partitioned, Frequency-Partitioned, and Block-Partitioned patch aggregation with $x$ and $y$ as the number of time and frequency chunks. The best and second-best results for each column in each layer fusion strategy are \textbf{bold} and \underline{underlined}, respectively.}}
\label{tab:results-ablation-strategy-1}
\begin{tabular}{ccclccccc}

\toprule
& & &\textbf{Dataset} & \multicolumn{2}{c}{\textbf{EMO}} & \textbf{GTZAN} & \multicolumn{2}{c}{\textbf{MTT}} \\ 
& & & \textbf{Type} & \multicolumn{2}{c}{\textit{Global}} & \textit{Global} & \multicolumn{2}{c}{\textit{Global}} \\
& & & \textbf{Task} & \multicolumn{2}{c}{Emotion} & Genre & \multicolumn{2}{c}{Tagging} \\
\midrule
\midrule
\textbf{Layer Fusion} & \textbf{Dim} & \textbf{Pooling} & \textbf{Patch Aggregation} & R2$^{\text{V}}$ ($\uparrow$) & R2$^{\text{A}}$ ($\uparrow$) & Acc ($\uparrow$) & ROC ($\uparrow$) & AP ($\uparrow$) \\
\midrule
\multirow{29}{*}{\makecell{\textbf{Weighted}\\ \textbf{Sum}}} & \multirow{4}{*}{8192} & \multirow{4}{*}{1} & Time-8 & \textbf{62.5} & 76.8 & 83.8 & 91.2 & 40.1 \\
& & & Freq-8 & 60.9 & 76.6 & 85.2 & 91.3 & 39.3 \\
& & & Block-2$\times$4 & 54.5 & 75.6 & 85.2 & 91.2 & 39.4 \\
& & & Block-4$\times$2 & 61.0 & 77.0 & \underline{86.2} & 91.0 & 38.3 \\
\cmidrule{2-9}
& \multirow{7}{*}{4096} & \multirow{3}{*}{1} & Time-4 & \underline{61.3} & 74.4 & 84.8 & 90.8 & 38.4 \\
& & & Freq-4 & 60.4 & 74.7 & \underline{86.2} & 91.3 & 39.0 \\
& & & Block-2$\times$2 & 60.8 & 76.5 & \textbf{86.9} & 91.3 & 39.1 \\
\cmidrule{3-9}
& & \multirow{4}{*}{2} & Time-8 & 60.1 & 77.4 & 82.1 & 91.1 & 39.3 \\
& & & Freq-8 & 57.8 & 77.1 & 85.2 & 91.1 & 39.2 \\ 
& & & Block-2$\times$4 & 59.9 & 77.0 & \underline{86.2} & 91.2 & 39.1 \\
& & & Block-4$\times$2 & 61.1 & 77.1 & 83.8 & 90.8 & 38.8 \\
\cmidrule{2-9}
& \multirow{10}{*}{2048} & \multirow{3}{*}{1} & Time-2 & 60.3 & 75.6 & 85.2 & 91.3 & 39.6 \\
& & & Freq-2 & 60.7 & 74.9 & 85.2 & 91.4 & 39.9 \\
\cmidrule{3-9}
& & \multirow{3}{*}{2} & Time-4 & 60.8 & 77.1 & 85.5 & 91.1 & 39.0 \\
& & & Freq-4 & 59.0 & 76.3 & 85.5 & 91.4 & 40.2 \\ 
& & & Block-2$\times$2 & 59.2 & 76.0 & 85.2 & 91.4 & 39.9 \\
\cmidrule{3-9}
& & \multirow{4}{*}{4} & Time-8 & 57.3 & 73.5 & 81.3 & 90.8 & 38.4 \\
& & & Freq-8 & 56.8 & 74.8 & 84.8 & 91.3 & 39.8 \\ 
& & & Block-2$\times$4 & 55.2 & \underline{77.6} & 83.8 & 91.5 & \underline{40.4} \\
& & & Block-4$\times$2 & 60.8 & \textbf{77.8} & 84.1 & 91.0 & 39.0 \\
\cmidrule{2-9}
& \multirow{7}{*}{1024} & \multirow{1}{*}{1} & Standard & 60.8 & 74.1 & 81.7 & 91.5 & 40.3 \\
\cmidrule{3-9}
& & \multirow{2}{*}{2} & Time-2 & \underline{61.3} & 76.8 & 85.5 & \underline{91.6} & 40.1 \\
& & & Freq-2 & 60.3 & 75.8 & 85.5 & \underline{91.6} & 40.2 \\
\cmidrule{3-9}
& & \multirow{3}{*}{4} & Time-4 & 56.0 & 74.8 & 83.4 & 91.2 & 39.5 \\
& & & Freq-4 & 58.0 & 76.2 & 84.1 & 91.5 & 40.1 \\ 
& & & Block-2$\times$2 & 56.8 & 75.4 & 82.4 & \textbf{91.7} & \textbf{40.8} \\
\midrule
\multirow{32}{*}{\makecell{\textbf{MLP}\\ \textbf{Reduce}}} & \multirow{10}{*}{2048} & \multirow{2}{*}{1} & Time-2 & 55.9 & 74.7 & 83.4 & 90.9 & 39.5 \\
& & & Freq-2 & 55.6 & 75.4 & 84.8 & 91.1 & 39.5 \\
\cmidrule{3-9}
& & \multirow{3}{*}{2} & Time-4 & 54.6 & 75.9 & 85.2 & 90.9 & 39.0 \\
& & & Freq-4 & 55.6 & 75.9 & 82.1 & 90.9 & 39.3 \\
& & & Block-2$\times$2 & 56.4 & 77.6 & 84.1 & \underline{91.3} & 39.7 \\
\cmidrule{3-9}
& & \multirow{4}{*}{4} & Time-8 & 58.7 & 77.5 & 82.4 & 90.6 & 38.8 \\
& & & Freq-8 & 56.5 & 78.6 & 83.4 & 90.8 & 39.1 \\
& & & Block-2$\times$4 & 51.4 & 78.7 & 82.1 & 91.2 & \underline{39.8} \\
& & & Block-4$\times$2 & 55.8 & 78.5 & 82.4 & 90.7 & 38.5 \\
\cmidrule{2-9}
& \multirow{11}{*}{1024} & 1 & Standard & 54.3 & 74.5 & 85.5 & \textbf{91.4} & 39.3 \\
\cmidrule{3-9}
& & \multirow{2}{*}{2} & Time-2 & 55.2 & 77.1 & \underline{86.6} & 90.8 & 38.7 \\
& & & Freq-2 & 55.3 & 76.6 & \textbf{86.9} & \underline{91.3} & \textbf{40.0} \\
\cmidrule{3-9}
& & \multirow{3}{*}{4} & Time-4 & 52.3 & 74.9 & 82.1 & 91.2 & 39.7 \\
& & & Freq-4 & 53.4 & 77.5 & 83.1 & 91.2 & 39.3 \\
& & & Block-2$\times$2 & 54.7 & 78.6 & 83.1 & 91.0 & 39.3 \\
\cmidrule{3-9}
& & \multirow{4}{*}{8} & Time-8 & 54.5 & 75.6 & 80.0 & 91.0 & 38.6 \\
& & & Freq-8 & 54.0 & 77.5 & 81.7 & 91.0 & 39.3 \\
& & & Block-2$\times$4 & 53.2 & 77.8 & 82.1 & \underline{91.3} & 39.6 \\
& & & Block-4$\times$2 & 57.0 & 77.7 & 82.8 & 91.1 & 39.5 \\
\cmidrule{2-9}
& \multirow{9}{*}{512} & 2 & Standard & 55.8 & 76.8 & \underline{86.6} & 91.2 & \underline{39.8} \\
\cmidrule{3-9}
& & \multirow{2}{*}{4} & Time-2 & 50.9 & 75.7 & 81.0 & 91.2 & 39.3 \\
& & & Freq-2 & 53.9 & 77.9 & 83.4 & \underline{91.3} & 39.4 \\
\cmidrule{3-9}
& & \multirow{3}{*}{8} & Time-4 & 54.1 & 76.1 & 81.4 & 90.9 & 38.6 \\
& & & Freq-4 & 56.6 & \underline{79.5} & 82.4 & \textbf{91.4} & 39.5 \\
& & & Block-2$\times$2 & 57.2 & \textbf{79.7} & 81.7 & 91.2 & 39.4 \\
\cmidrule{3-9}
& & \multirow{2}{*}{16} & Time-8 & \textbf{63.4} & 72.6 & 77.2 & 91.0 & 39.0 \\
& & & Freq-8 & \underline{59.3} & 76.2 & 79.7 & 91.2 & 39.7 \\
\bottomrule

\end{tabular}
\end{center}
\end{table*}

\begin{table*}[p]
\begin{center}
\caption{\revision{Ablation study of different layer fusion and patch aggregation strategies with various pooling factors on PupuM2D-Large in MARBLEv2's global tasks (2/2). ``Standard'', ``Time-$x$'', ``Freq-$y$'', and ``Block-$x$$\times$$y$'' mean Standard, Time-Partitioned, Frequency-Partitioned, and Block-Partitioned patch aggregation with $x$ and $y$ as the number of time and frequency chunks. The best and second-best results for each column in each layer fusion strategy are \textbf{bold} and \underline{underlined}, respectively.}}
\label{tab:results-ablation-strategy-2}
\begin{tabular}{ccclcccccccc}

\toprule
& & &\textbf{Dataset} & \multicolumn{2}{c}{\textbf{MTG}} & \multicolumn{2}{c}{\textbf{MTG}} & \multicolumn{2}{c}{\textbf{MTG}} & \multicolumn{2}{c}{\textbf{MTG}} \\ 
& & & \textbf{Type} & \multicolumn{2}{c}{\textit{Global}} & \multicolumn{2}{c}{\textit{Global}} & \multicolumn{2}{c}{\textit{Global}} & \multicolumn{2}{c}{\textit{Global}} \\
& & & \textbf{Task} & \multicolumn{2}{c}{Instrument} & \multicolumn{2}{c}{MoodTheme} & \multicolumn{2}{c}{Genre} & \multicolumn{2}{c}{Top50} \\
\midrule
\midrule
\textbf{Layer Fusion} & \textbf{Dim} & \textbf{Pooling} & \textbf{Patch Aggregation} & ROC ($\uparrow$) & AP ($\uparrow$) & ROC ($\uparrow$) & AP ($\uparrow$) & ROC ($\uparrow$) & AP ($\uparrow$) & ROC ($\uparrow$) & AP ($\uparrow$) \\
\midrule
\multirow{29}{*}{\makecell{\textbf{Weighted}\\ \textbf{Sum}}} & \multirow{4}{*}{8192} & \multirow{4}{*}{1} & Time-8 & 76.4 & 19.4 & \textbf{75.1} & \textbf{14.8} & 84.3 & \underline{18.7} & \underline{81.5} & \textbf{29.3} \\
& & & Freq-8 & 75.1 & 18.9 & 73.7 & 13.9 & 83.0 & 17.5 & 80.6 & 28.0 \\
& & & Block-2$\times$4 & 75.5 & 19.5 & 74.0 & 14.0 & 83.1 & 17.7 & 80.8 & 28.3 \\
& & & Block-4$\times$2 & 76.2 & 19.7 & 74.6 & 14.4 & 83.7 & 17.9 & 81.2 & 28.5 \\
\cmidrule{2-12}
& \multirow{7}{*}{4096} & \multirow{3}{*}{1} & Time-4 & \textbf{76.6} & \underline{19.9} & 74.9 & 14.2 & 83.9 & 18.6 & 81.3 & 28.9 \\
& & & Freq-4 & 76.0 & \textbf{20.0} & 74.2 & 13.9 & 83.3 & 17.8 & 80.8 & 28.1 \\
& & & Block-2$\times$2 & 76.1 & 19.6 & 74.3 & 13.7 & 83.6 & 18.1 & 81.1 & 29.0 \\
\cmidrule{3-12}
& & \multirow{4}{*}{2} & Time-8 & 76.0 & 19.2 & 74.3 & 14.1 & \textbf{85.6} & \textbf{18.9} & \textbf{81.6} & \underline{29.1} \\
& & & Freq-8 & 75.4 & 19.7 & 73.6 & 13.4 & 82.8 & 17.4 & 80.7 & 28.0 \\
& & & Block-2$\times$4 & 75.2 & 19.4 & 74.1 & 13.8 & 83.9 & 18.0 & 80.8 & 28.2 \\
& & & Block-4$\times$2 & \underline{76.5} & 19.7 & \underline{75.0} & 14.2 & 84.0 & 18.0 & 81.3 & 28.7 \\
\cmidrule{2-12}
& \multirow{10}{*}{2048} & \multirow{3}{*}{1} & Time-2 & 75.4 & 19.4 & 73.6 & 13.8 & 83.9 & 18.5 & \textbf{81.6} & 28.8 \\
& & & Freq-2 & 75.5 & 19.7 & 73.7 & 14.0 & 83.8 & 18.3 & 80.7 & 28.4 \\
\cmidrule{3-12}
& & \multirow{3}{*}{2} & Time-4 & 75.4 & 19.5 & 74.7 & 14.3 & 84.2 & 18.6 & \textbf{81.6} & 28.8 \\
& & & Freq-4 & 75.4 & 19.0 & 74.0 & 14.2 & 83.4 & 18.0 & 80.8 & 28.3 \\ 
& & & Block-2$\times$2 & 75.4 & 19.4 & 74.4 & 14.2 & 83.9 & 18.1 & 81.1 & 28.6 \\
\cmidrule{3-12}
& & \multirow{4}{*}{4} & Time-8 & 75.6 & 19.4 & 74.1 & 13.6 & \underline{84.9} & 18.5 & 81.3 & \textbf{29.3} \\
& & & Freq-8 & 75.0 & 18.8 & 73.2 & 13.6 & 83.4 & 17.6 & 80.8 & 28.1 \\ 
& & & Block-2$\times$4 & 75.3 & 19.2 & 74.4 & 14.1 & 84.1 & 18.1 & 81.1 & 28.7 \\
& & & Block-4$\times$2 & 75.6 & 19.5 & \underline{75.0} & \underline{14.5} & 84.4 & 18.5 & \underline{81.5} & 28.9 \\
\cmidrule{2-12}
& \multirow{7}{*}{1024} & \multirow{1}{*}{1} & Standard & 75.0 & 19.1 & 74.0 & 14.0 & 83.2 & 18.2 & 80.9 & 28.4 \\
\cmidrule{3-12}
& & \multirow{2}{*}{2} & Time-2 & 75.4 & 19.7 & 74.4 & 14.2 & 83.9 & 18.2 & 81.3 & 29.0 \\
& & & Freq-2 & 75.1 & 19.3 & 74.7 & 14.3 & 83.6 & 18.0 & 81.0 & 28.7 \\
\cmidrule{3-12}
& & \multirow{4}{*}{3} & Time-4 & 75.4 & 19.3 & 74.6 & 14.4 & 84.0 & 18.5 & \textbf{81.6} & 29.0 \\
& & & Freq-4 & 75.1 & 19.4 & 74.2 & 13.8 & 83.3 & 17.9 & 80.8 & 28.4 \\ 
& & & Block-2$\times$2 & 75.3 & 19.3 & 74.6 & 14.2 & 83.9 & 18.2 & 80.9 & 28.7 \\
\midrule
\multirow{32}{*}{\makecell{\textbf{MLP}\\ \textbf{Reduce}}} & \multirow{10}{*}{2048} & \multirow{2}{*}{1} & Time-2 & 77.5 & \underline{20.9} & \underline{76.1} & \underline{15.2} & \underline{85.6} & \underline{20.0} & 82.3 & \textbf{30.5} \\
& & & Freq-2 & 76.9 & 20.2 & 73.9 & 14.1 & 85.2 & 19.5 & 81.4 & 29.7 \\
\cmidrule{3-12}
& & \multirow{3}{*}{2} & Time-4 & \textbf{78.4} & \textbf{21.2} & 75.0 & 14.4 & \underline{85.6} & 19.8 & 82.5 & \underline{30.4} \\
& & & Freq-4 & 76.1 & 20.2 & 74.5 & 14.6 & 84.8 & 18.7 & 81.7 & 30.0 \\
& & & Block-2$\times$2 & 77.6 & 20.8 & 74.9 & 14.9 & 85.2 & 19.0 & 82.5 & \textbf{30.5} \\
\cmidrule{3-12}
& & \multirow{4}{*}{4} & Time-8 & 76.0 & 20.8 & \underline{76.1} & 15.0 & \textbf{86.1} & \textbf{20.1} & \textbf{82.8} & \textbf{30.5} \\
& & & Freq-8 & 76.0 & 19.3 & 74.1 & 14.2 & 85.5 & 19.3 & 81.6 & 29.4 \\
& & & Block-2$\times$4 & 76.9 & 20.7 & 74.9 & 14.3 & 84.8 & 19.1 & 82.0 & 30.2 \\
& & & Block-4$\times$2 & 76.7 & 20.8 & 75.3 & 14.7 & \underline{85.6} & 19.7 & 81.7 & 29.9 \\
\cmidrule{2-12}
& \multirow{11}{*}{1024} & 1 & Standard & 75.2 & 20.3 & 74.9 & 14.2 & 85.1 & 19.8 & 82.1 & 29.9 \\
\cmidrule{3-12}
& & \multirow{2}{*}{2} & Time-2 & \underline{77.7} & 20.8 & 75.3 & 14.9 & 85.3 & 19.7 & 82.1 & 30.1 \\
& & & Freq-2 & 76.2 & 20.1 & 75.1 & 14.8 & 85.0 & 19.3 & 81.6 & 30.0 \\
\cmidrule{3-12}
& & \multirow{3}{*}{4} & Time-4 & 76.3 & 20.4 & 75.4 & 14.5 & 85.4 & 19.7 & 82.4 & \underline{30.4} \\
& & & Freq-4 & 76.2 & 20.6 & 74.3 & 14.1 & 84.4 & 18.8 & 81.6 & 29.7 \\
& & & Block-2$\times$2 & 76.3 & 20.5 & 75.2 & 14.8 & 85.5 & 19.6 & 81.8 & 30.2 \\
\cmidrule{3-12}
& & \multirow{4}{*}{8} & Time-8 & 77.5 & 20.3 & \textbf{76.2} & \textbf{15.3} & 85.5 & \underline{20.0} & 82.3 & 30.3 \\
& & & Freq-8 & 75.9 & 19.4 & 74.6 & 14.0 & 85.2 & 18.7 & 81.4 & 28.9 \\
& & & Block-2$\times$4 & 76.5 & 20.4 & 74.6 & 13.8 & 85.3 & 19.3 & 81.9 & 29.8 \\
& & & Block-4$\times$2 & 76.2 & 20.5 & 75.0 & 14.2 & 85.3 & 19.4 & 82.0 & 30.0 \\
\cmidrule{2-12}
& \multirow{9}{*}{512} & 2 & Standard & 75.6 & 19.8 & 75.2 & 14.2 & 84.9 & 18.9 & 81.8 & 29.9 \\
\cmidrule{3-12}
& & \multirow{2}{*}{4} & Time-2 & 74.9 & 19.9 & 75.5 & 14.6 & 85.0 & 19.5 & 82.0 & 29.9 \\
& & & Freq-2 & 74.9 & 20.0 & 75.0 & 14.6 & 84.5 & 18.8 & 81.5 & 29.6 \\
\cmidrule{3-12}
& & \multirow{3}{*}{8} & Time-4 & 75.7 & 19.8 & 75.9 & 14.9 & 84.7 & 19.5 & 82.6 & 30.3 \\
& & & Freq-4 & 75.0 & 20.0 & 74.6 & 14.8 & 84.6 & 18.4 & 81.8 & 29.7 \\
& & & Block-2$\times$2 & 75.8 & 20.2 & 75.4 & 14.4 & 84.7 & 19.1 & 82.2 & 29.8 \\
\cmidrule{3-12}
& & \multirow{2}{*}{16} & Time-8 & 76.1 & 20.1 & 75.9 & 14.7 & 85.3 & 19.5 & \underline{82.7} & 29.8 \\
& & & Freq-8 & 75.4 & 19.8 & 74.7 & 14.2 & 84.1 & 18.5 & 81.8 & 29.2 \\
\bottomrule

\end{tabular}
\end{center}
\end{table*}

\begin{table*}[t]
\begin{center}
\caption{\revision{Ablation study of different layer fusion strategies with various pooling factors on PupuM2D-Large in MARBLEv2 benchmark's local tasks. The best and second-best results for each column in each strategy are \textbf{bold} and \underline{underlined}, respectively.}}
\label{tab:results-ablation-strategy-3}
\begin{tabular}{cclcccc}

\toprule
& & \textbf{Dataset} & \textbf{GS} & \textbf{GTZAN} & \multicolumn{2}{c}{\textbf{HookTheory}} \\ 
& & \textbf{Type} & \textit{Local} & \textit{Local} & \textit{Local} & \textit{Local} \\
& & \textbf{Task} & Key & Rhythm & Key & Structure \\
\midrule
\midrule
\textbf{Layer Fusion} & \textbf{Dim} & \textbf{Pooling} & Acc$^{\text{Refined}}$ ($\uparrow$) & F1$^{\text{Beat}}$ ($\uparrow$) & Acc$^{\text{Refined}}$ ($\uparrow$) & Acc ($\uparrow$) \\
\midrule
\multirow{3}{*}{\makecell{\textbf{Weighted}\\ \textbf{Sum}}} & 8192 & 1 & \textbf{66.1} & \textbf{91.0} & \textbf{72.9} & \textbf{57.6} \\
& 4096 & 2 & 62.9 & \underline{90.6} & 71.4 & \underline{56.4} \\
& 2048 & 4 & \underline{64.2} & 90.5 & \underline{71.7} & \textbf{57.6} \\
\midrule
\multirow{3}{*}{\makecell{\textbf{MLP}\\ \textbf{Reduce}}} & 2048 & 4 & \textbf{63.8} & \textbf{90.5} & \textbf{72.9} & \textbf{57.2} \\
& 1024 & 8 & \underline{62.4} & \underline{90.4} & \underline{71.8} & 56.4 \\
& 512 & 16 & 60.1 & \textbf{90.5} & 68.4 & \underline{56.8} \\
\bottomrule

\end{tabular}
\end{center}
\vspace{-16pt}
\end{table*}

\subsubsection{Emotional Analysis}
\label{subsubsec:task-emotion}

Emotional analysis is a \textit{global task} that aims to predict the emotional arousal and valence values from music signals. We use the EmoMusic (EMO)~\cite{emomusic} dataset for evaluation and report the determination coefficients for valence (R$^2_V$) and arousal (R$^2_A$) scores as metrics.

\subsubsection{Instrument Classification}
\label{subsubsec:task-instrument}

Instrument classification is a \textit{global task} that aims to identify the presence of specific musical instruments within a given music track. We use the MTG~\cite{mtg} dataset for evaluation and report the macro-averaged ROC-AUC and AP scores as the metrics.

\subsubsection{Key Detection}
\label{subsubsec:task-key}

Key detection is a \textit{local task} that aims to estimate the frame-level musical key of a given track. We use the GiantSteps (GS)~\cite{giantsteps} and HookTheory~\cite{hooktheory} datasets for evaluation and report the refined accuracy~\cite{mireval} as the metric.

\subsubsection{Beat Tracking}
\label{subsubsec:task-beat}

Beat tracking is a \textit{local task} that aims to locate the frame-level beat timestamps (Beats, Downbeats, and Tempo) of music tracks. We use the GTZAN~\cite{gtzan} dataset for evaluation and report the F1-score (F1$^{\text{Beat}}$) as the metric.

\subsubsection{Music Structure Analysis}
\label{subsubsec:task-structure}

Music structure analysis is a \textit{local task} that aims to categorize each frame into distinct functional sections, such as Intro, Verse, Chorus, Bridge, and Outro. We use the HookTheory~\cite{hooktheory} dataset for evaluation and report the frame-level accuracy as the metric.

\subsubsection{Music Tagging}
\label{subsubsec:task-tagging}

Music tagging is a \textit{global task} that aims to assign descriptive tags to different music clips. We use the MagnaTagATune (MTT)~\cite{MTT} and MTG~\cite{mtg} datasets, each with the top 50 most frequent tags for evaluation. We report the macro-averaged ROC-AUC and AP scores as the metrics.

\subsubsection{Genre Classification}
\label{subsubsec:task-genre}

Genre classification is a \textit{global task} that categorizes music tracks into distinct genres. We use the GTZAN~\cite{gtzan} and MTG-Jamendo (MTG)~\cite{mtg} datasets for evaluation. For metrics, we report the standard accuracy on the GTZAN dataset. We report the Receiver Operating Characteristic Area Under the Curve (ROC-AUC) and Average Precision (AP) on the multi-label MTG dataset.

\subsection{Experimental Results}

\subsubsection{Comparison with Baseline Models}

Tables~\ref{tab:results-marble-1} and~\ref{tab:results-marble-2} illustrate the evaluation results on the MARBLEv2 benchmark. Notably, across all 2D models, including both our PupuM2D variants and the reproduced audio baselines, the reported metrics are the optimal results obtained by sweeping across the various inference paradigms to ensure a fair comparison. \revision{It is also worth noting that the 2D audio SSL baselines we adapted directly from their official releases hold inductive bias from ViTs in the evaluation results, where near-square patches (such as $16\times16$) will facilitate the representation quality and result in higher scores. However, we would like to highlight that this design inevitably incurs a severe trade-off, where the resulting small frame rate will make these models incapable of performing fine-grained local tasks.} Overall, it can be shown that the proposed PupuM2D model has exceptional performance, achieving SOTA results across most tasks.

\textbf{Superiority over 1D Sequence-Based Baselines:}
It can be observed that the 2D spectrogram-based models generally outperform standard 1D sequence-based models across a wide range of tasks, \revision{including the baselines that we directly adapted from their official release, which are trained with audio events data only}. This validates the inherent advantage of preserving structural time-frequency information over treating the audio as 1D sequences. Such a superiority is particularly effective in local tasks where 2D spectrograms naturally provide intuitive visual cues. For instance, the 5M parameter PupuM2D-Tiny achieves 64.2\% on GS Key, outperforming all baseline models.

\textbf{Effectiveness against 2D Audio Baselines:}
Compared to general 2D audio SSL baselines, PupuM2D demonstrates clear superiority on tasks that require high-level abstract understanding, mainly including emotion recognition, genre classification, and instrument classification. \revision{Compared to 2D baselines that we reproduced, PupuM2D is still better at all fine-grained local tasks}. These observations confirm the effectiveness of our domain-specific modifications on model architecture, training scheme, and inference diagrams.

\textbf{Robustness across Global and Local Tasks:}
For fine-grained local tasks, PupuM2D shows clear advantages in GS key detection and GTZAN beat tracking. However, its performance on HookTheory structure analysis is mostly on par with the baselines. This reflects an inherent trade-off when applying linear probing to 2D models for local tasks: applying pooling inevitably destroys localized time-frequency structure, whereas completely bypassing pooling results in an exceptionally high feature dimension, making the linear probe harder to optimize. Conversely, for global tasks, PupuM2D demonstrates consistent performance gains from our proposed patch aggregation strategies, especially in emotion regression and various MTG global tasks (Instrument and TOP50).

\textbf{Scaling Behavior:}
Consistent parameter scaling behavior can be observed across the PupuM2D variants. Downstream task performance exhibits a clear and steady upward trajectory as model capacity increases from the Tiny (5M) and Small (22M) variants, culminating at the PupuM2D-Large (307M) scale. Across both global and local tasks, the Large variant consistently emerges as the optimal configuration. Scaling further to PupuM2D-Huge (632M) yields diminishing returns, with slight performance regressions observed across several linear probing metrics. This saturation is a well-documented phenomenon in SSL under linear probing evaluations: as model capacity becomes excessively large, the representations often form highly complex, non-linear structures. While they may have deeper information, they become inherently more difficult for a simple single-layer linear classifier to learn~\cite{mae, ijepa, dino, dino2, dino3}. As a result, PupuM2D-Large strikes the ideal balance between model capacity and probing compatibility.

\subsubsection{Ablation Study on Model Architecture}

We ablated all modern ViT performance enhancements and domain-specific modifications on PupuM2D-Large to systematically show their individual contributions, as illustrated in Tables~\ref{tab:results-ablation-architecture-1} and \ref{tab:results-ablation-architecture-2}. We found that these modifications can be categorized into two groups: \textit{performance enhancers} that improve the performance, and \textit{stability guardians} that avoid representation collapse.

\textbf{Performance Enhancers:} Removing \textit{SwiGLU}, \textit{QKNorm}, and \textit{Mixing Masking Strategy} leads to noticeable performance drops across nearly all metrics. This indicates that \textit{SwiGLU} effectively expands the ViT representation capacity, \textit{QKNorm} stabilizes the magnitudes of attention scores for high-variance spectrogram features, and incorporating blockwise and time-frequency masking forces the model to capture longer-range temporal and wider-spectral contextual dependencies.

\textbf{Stability Guardians:} Notably, several standard components are incompatible with PupuM2D. Our ablation study illustrates that substituting the \textit{Smoothed $L_1$ Loss} with standard \textit{$L_1$ Loss}, or equipping ViT with standard \textit{DropPath}, \textit{LayerScale}, and \textit{Batch Normalization}, will result in training instabilities and representation collapse. Meanwhile, feeding the full input sequence to the target encoder, as done in standard JEPA implementations~\cite{ijepa}, also leads to representation collapse, confirming the shortcut learning hypothesis in M2D~\cite{m2d}.

\begin{figure*}[t]
    \centering
    \includegraphics[width=0.95\textwidth]{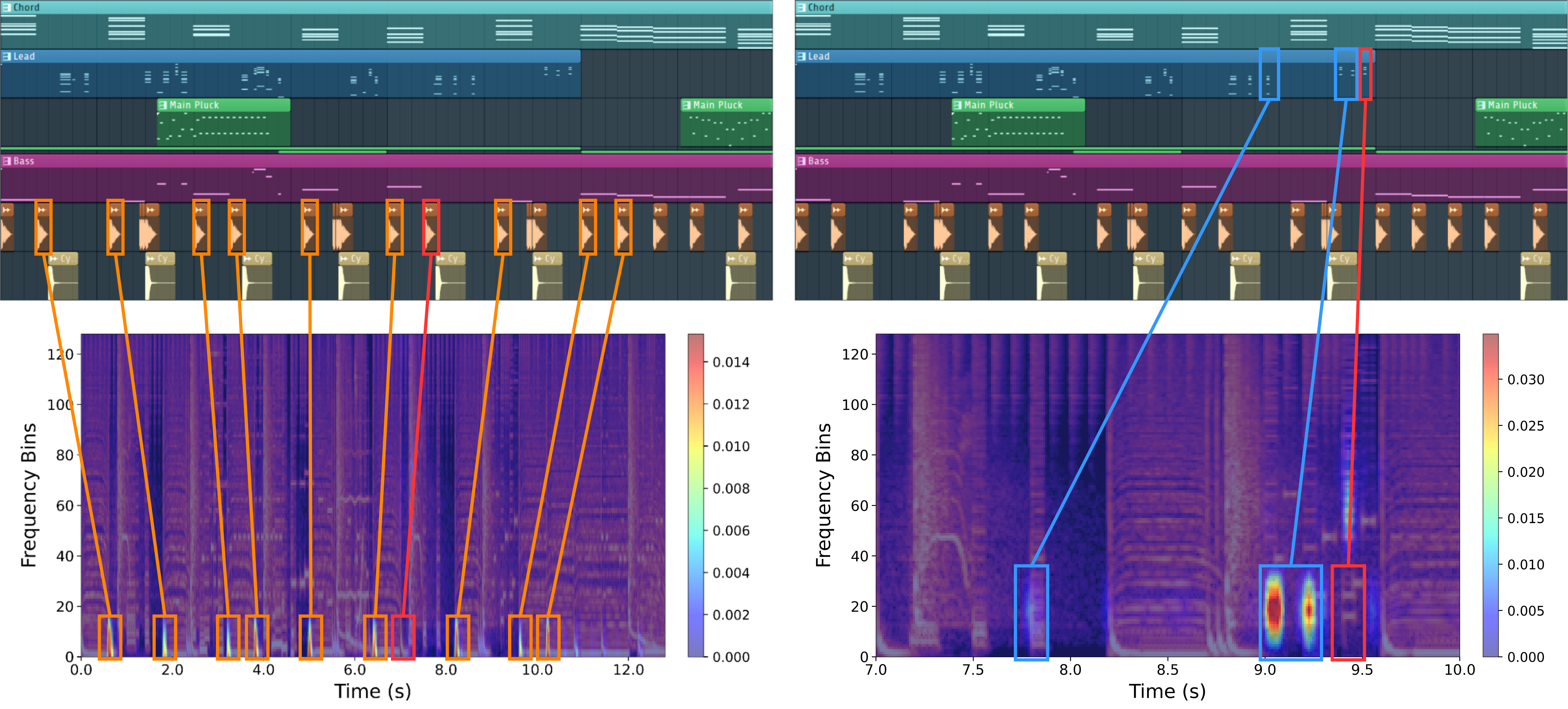}
    \caption{\revision{Illustration of PupuM2D's predictor attention maps. The top panels display a multitrack project in a DAW, while the bottom panels show the corresponding 2D spectrograms overlaid with the attention heatmaps. Note that the color bars displayed on the right of the bottom panels denote the attention weights rather than the energy of the spectrograms.
    The red bounding boxes denote the masked target regions that the model aims to predict, while the other colored bounding boxes map individual track components to their corresponding time-frequency patterns in the 2D spectrogram, specifically highlighting the regions that has high attention weights. In particular, the bottom-left panel shows the attention map when predicting a masked Kick drum, whereas the bottom-right panel displays the attention map when predicting the masked root notes of a Lead sound.}}
    \label{fig:case}
    \vspace{-20pt}
\end{figure*}

\subsubsection{Ablation Study on Inference Paradigms}
\label{sec:ablation}

To validate our proposed 2D inference paradigms, we conducted extensive ablation studies on the layer fusion and patch aggregation strategies with various pooling factors using the PupuM2D-Large model. The results are in Tables~\ref{tab:results-ablation-strategy-1}, \ref{tab:results-ablation-strategy-2}, and \ref{tab:results-ablation-strategy-3}.

\textbf{Impact of Pooling on Local Tasks:} 
For fine-grained local tasks, it can be observed that applying patch-wise pooling along the frequency dimension significantly degrades the quality of spectral structure, which is vital for precise harmonic and temporal understanding. As shown in Table~\ref{tab:results-ablation-strategy-3}, the \textit{Weighted Sum} strategy without any pooling consistently achieves the highest performance across all evaluated local tasks, demonstrating the necessity of preserving uncompressed spectral dimensionality for fine-grained frame-level analysis.

\textbf{Effectiveness of Partitioned Patch Aggregation:} 
Among the different patch aggregation methods for global tasks, \textit{Time-Partitioned} and \textit{Block-Partitioned Patch Aggregation} consistently outperform the \textit{Standard GAP} and \textit{Frequency-Partitioned Patch Aggregation}, illustrating that preserving coarse temporal sequences and spatial correlations is more effective than flattening the spectrogram. We also observe that the optimal partition parameters depend on the specific task, as different MIR tasks rely on events at different resolutions.

\textbf{Layer Fusion versus Dataset Scale:} 
For global tasks, the optimal layer fusion strategy strongly correlates with the scale of the downstream dataset. On relatively smaller datasets (such as EMO, GTZAN, and MTT), the \textit{Weighted Sum} approach generally outperforms \textit{MLP Reduce}. Because \textit{Weighted Sum} directly aggregates layers via simple learnable weights, it effectively prevents overfitting when downstream data is limited. Conversely, on the massive MTG benchmark datasets (Table~\ref{tab:results-ablation-strategy-2}), \textit{MLP Reduce} coupled with appropriate pooling yields better performance, as the abundance of training data avoids the previous overfitting problem and thus allows the MLP head to robustly learn complex, non-linear reductions from the compressed multi-layer 2D patches.

\subsubsection{Attention Map Visualization}

We additionally conduct qualitative case studies to investigate whether PupuM2D learns musically meaningful patterns that inherently align with the intuition of human music producers. 
In particular, we extract the queries and keys from the last transformer layer of the predictor and then inject the 2D rotary position embeddings. 
After that, we compute the scaled dot-product attention and apply a softmax function to obtain the probability distribution. 
The resulting attention weights are subsequently averaged across all attention heads. 
The attention heatmap is upsampled to the original resolution using bilinear interpolation and overlaid on the original Mel-spectrogram for visualization. 
The maximum color intensity is capped at the 99.9th percentile to ensure visual clarity. The whole process can be shown as:
\begin{equation}
\mathbf{A} = \mathcal{F}_{\text{up}} \left( \frac{1}{N_{\text{h}}} \sum_{n=1}^{N_{\text{h}}} \mathrm{Softmax} \left( \frac{ \mathrm{RoPE}(\mathbf{q}_{n}) \cdot \mathrm{RoPE}(\mathbf{k}_{n})^\top}{\sqrt{d_k}} \right) \right),
\end{equation}

where $\mathbf{A}$ is the attention heatmap, $\mathcal{F}_{\text{up}}(\cdot)$ is the upsampling function (bilinear interpolation), $N_{\text{h}}$ is the number of attention heads, $\mathbf{q}_{n}$ is the query vector of the target patch for the attention head $n$, $\mathbf{k}_{n}$ is the key matrix of all context tokens for the attention head $n$, $\mathrm{RoPE}(\cdot)$ denotes the injection of 2D rotary position embeddings, and $d_k$ is the scaling factor.

We then intuitively align the attention heatmaps with the multi-track project for analysis, as illustrated in Figure~\ref{fig:case}. 
As shown in the bottom-left panel, when predicting a masked Kick drum, the model does not simply look at the surrounding pixels; instead, its attention spans broadly across the music segment, selectively highlighting other Kick drum distributions with similar rhythm patterns, demonstrating its ability to capture long-term dependencies. Conversely, as shown in the bottom-right panel, when the model attempts to predict the masked root notes of a Lead sound, its attention spans across the remaining unmasked higher harmonics and the adjacent melodic Lead progressions just before the masked region, illustrating its ability to understand musical context.

Such visual evidence strongly validates our core intuition. In MIDI-based music production, music is naturally represented as 2D time-frequency grids, which closely correspond to 2D spectrograms. Just like music producers can ``reconstruct'' a multi-track project from a spectrogram, the PupuM2D, pre-trained directly on music spectrograms, also develops this capability. As shown in Figure~\ref{fig:case}, PupuM2D naturally enables zero-shot disentanglement of individual tracks and long-range musical structural analysis, thereby making many MIR tasks trivial to probe and achieving the SOTA performance.

\section{Conclusion}

This paper proposes PupuM2D, a musical M2D explicitly tailored for 2D self-supervised music representation learning. 
To address the inherent limitations of standard 1D sequence-based musical SSL models, which discard rich spatial and harmonic structures in the time-frequency domain, PupuM2D directly models time-frequency representations by predicting the latent embeddings of masked 2D spectrogram patches. 
To optimize this paradigm for the music domain, we made critical architectural modifications to ensure training stability and proposed novel inference paradigms to improve downstream probing performance. 
Experimental results on the MARBLEv2 benchmark show that PupuM2D consistently outperforms SOTA 1D sequence-based SSL models and 2D spectrogram-based audio baselines across various MIR tasks. 
Finally, our case study on attention maps shows that PupuM2D can capture musically meaningful patterns within the 2D time-frequency domain, confirming our intuition and its effectiveness.

\bibliographystyle{IEEEtran}
\bibliography{ref}

\end{document}